\documentstyle[seceq,preprint]{ptptex}

\catcode`@=11     

\def\partialslash{\rlap{\hbox{/}}\partial}
\def\Aslash{\rlap{\hbox{/}} A}

\def\eqalign#1{\null\,\vcenter{\openup\jot\m@th
  \ialign{\strut\hfil$\displaystyle{##}$&$\displaystyle{{}##}$\hfil
      \crcr#1\crcr}}\,}

\def\meqalign#1{\null\,\vcenter{\openup\jot\m@th
  \ialign{\strut\hfil$\displaystyle{##}$&&$\displaystyle{{}##}$\hfil
      \crcr#1\crcr}}\,}

\catcode`@=12   

\def\beq{\begin{equation}} 
\def\eeq{\end{equation}}
\def\be{\beq}
\def\ee{\eeq} 

\def\partialslash{\rlap{\hbox{/}}\partial}
\def\Aslash{\rlap{\hbox{/}} A}

\def\PPh#1{\setbox0\hbox{$#1\rm I$}\mathord{\vcenter{\ialign{$#1\rm##$\cr
I\cr\noalign{\nointerlineskip \vskip-0.541\ht0}P\cr}}}}

\def\Ph{{\mathpalette\PPh{}}}
\def\Ed{{\mathord{\mkern5mu\mathaccent"7020{\mkern-5mu\partial}}}}

\title{
Teukolsky Master Equation: De Rham wave equation \\
for the gravitational and electromagnetic fields in vacuum}

\author{
Donato {\sc Bini}$^{*,\dag}$, 
Christian {\sc Cherubini}$^{\sharp,\dag}$, 
Robert T. {\sc Jantzen}$^{\P,\dag}$, 
\\
and Remo {\sc Ruffini}$^{\flat,\dag}$ 
}
\inst{
$^{*}$\,  Istituto per le Applicazioni del Calcolo, ``M. Picone,'' C.N.R., 
  \\I-- 00161 Rome, Italy. \\
$^{\dag}$\,  I.C.R.A., 
  Univ. of Rome, I--00185 Roma, Italy.\\
$^{\P}$\, Dept. of Math. Sciences, Villanova Univ., 
  Villanova, PA 19085, USA. \\
$^{\sharp}$\, Dip. di Fisica ``E.R.
Caianiello,'' Univ. di Salerno, I--84081, Italy.\\
$^{\flat}$\, Physics Dept., 
  Univ. of Rome, I--00185 Roma, Italy.
}

\recdate{
}

\abst{
A new version of the Teukolksy Master Equation, describing any massless field of different spin $s=1/2,1,3/2,2$ in the Kerr black hole, is presented here in the form of a wave equation containing additional curvature terms. 
These results suggest a relation 
between curvature perturbation theory in general relativity and the exact wave equations satisfied by the Weyl and the Maxwell tensors, known in the literature as the de Rham-Lichnerowicz Laplacian equations.
We discuss these Laplacians both in the Newman-Penrose formalism and in the Geroch-Held-Penrose variant for an arbitrary vacuum spacetime.
Perturbative expansion of these wave equations 
results in a recursive scheme valid for higher orders.
This approach, apart from the obvious implications for the gravitational and electromagnetic wave propagation on a curved spacetime, explains and extends the results in the literature for perturbative analysis
by clarifying their true origins in the exact theory. 
}

\begin{document}

\maketitle

\section{Introduction}

In recent years there has been considerable interest in perturbations of the gravitational field and in other spin $s$ massless test fields ($s=1/2,1,3/2,2$) imposed on a gravitational background (especially on a black hole background, which is of particular interest for studying features of gravitational collapse and consequent gravitational wave emission).
The approaches to perturbations existing in the literature can be divided essentially into two classes:
1) metric perturbations and 2) curvature perturbations.

Pioneering work in the first class was done by Regge, Wheeler and Zerilli, who produced important results 
for spherically symmetric black holes\cite{ReggeWheeler,Zerilli,Johnston} by decomposing the metric perturbations into multipoles which satisfy a set of (linearized) equations 
\cite{Lichnerowicz1,SchmidtDamour,Christodoulou,Sciama,Marsden,Choquet}. 
However, a) perturbing the metric components written with respect to some coordinate system is not a ``gauge invariant" procedure as pointed out by Stewart and Walker \cite{StewartWalker} and b) apart from spherically symmetric background solutions, this approach does not seem to be very useful. 

Working in the second class, i.e., perturbing directly the components of the curvature tensor, one can use frames adapted to the null principal directions of the Weyl tensor of the background spacetime: this allows special simplifications (even  partial decoupling) of the equations themselves.
However, this approach has problems too: perturbing the components of the curvature tensor with respect to some special adapted frame is not a ``tetrad invariant" procedure and one must add this difficulty to that of the gauge invariance which is still present.

The ``state of the art" regarding curvature perturbations is still represented by the work of Teukolsky \cite{Teukolsky} (in the context of the Newman-Penrose formalism\cite{NP}, NP in the following), which partially received its mathematical foundation by Stewart and Walker \cite{StewartWalker} and was subjected to some extensions by other authors \cite{Cohen1,Cohen2,Chrzanowski1,Chrzanowski2,Wald,Finley,Fayos,GHP}.
Very recently,  a gauge-invariant second-order perturbation study for a type D vacuum geometry was initiated by Lousto and Campanelli\cite{LoustoCampanelli}, also stimulated by relevant results concerning the second order perturbations of a Schwarzschild black hole\cite{Pullin}.

We introduce a new form of the Teukolksy Master Equation, closer to a ``generalized Klein Gordon" equation. This allows us to investigate exact wave equations on a general vacuum spacetime: the de Rham-Lichnerowicz equation satisfied by the Riemann tensor as well as the ordinary de Rham equation satisfied by the Maxwell tensor in a test field approximation.
All the results are formulated for tensor components with respect to any frame and then specialized to the Newman-Penrose formalism and to its further extension, the GHP formalism, introduced by Geroch, Held and Penrose \cite{GHP}.
Starting from the exact equations for the Riemann and Maxwell tensor components, 
it is easy to give a unified and complete picture of the existing theory of perturbations to any order, 
framing all the known results (which are scattered in many different formalisms) in a more general form.
In type D spacetimes all the known results, including those of Teukolsky\cite{Teukolsky}, Stewart and Walker\cite{StewartWalker} and Lousto and Campanelli\cite{LoustoCampanelli} 
are easily recovered and discussed. 
Finally, some remarks are given about how this approach can be generalized to half-integer spin test fields as well as to the case of nonvacuum spacetimes.

\section{The Newman-Penrose formalism: a brief review}

In this article we introduce the Newman-Penrose form of the generalized de Rham wave operator for the Weyl, Riemann and Maxwell tensors.
To accomplish this, it is first necessary to specify all the formalism conventions: 
we will follow exactly the notations of Chandrasekhar\cite{chandra}. 
For the sake of completeness we recall here some details.  

A Newman-Penrose frame is defined by four complex 
null vector fields 
${\bf e}_1={\bf l}$, ${\bf e}_2={\bf n}$, ${\bf e}_3={\bf m}$, 
${\bf e}_4={\bf m^*}$, 
satisfying the relations 
\begin{eqnarray}
\meqalign{
&{\bf l} \cdot {\bf m} ={\bf l} \cdot {\bf m^*} ={\bf n} \cdot {\bf m} ={\bf n} \cdot {\bf m^*} =0, \cr
&{\bf l} \cdot {\bf l} ={\bf n} \cdot {\bf n} ={\bf m} \cdot {\bf m} ={\bf m^*} \cdot {\bf m^*} =0, \cr
&{\bf l} \cdot {\bf n} =1,\,\,\,\,\,\,\, {\bf m} \cdot {\bf m^*} =-1,\,\,\,}
\end{eqnarray}
so that the components of the metric tensor in this frame are 
\beq \eta_{ab}=\eta^{ab}=
\pmatrix{0&1&0&0\cr
           \noalign{\smallskip}
           1&0&0&0\cr
\noalign{\smallskip}
           0&0&0&-1\cr
\noalign{\smallskip}
           0&0&-1&0\cr} \ .\label{ETANP}
\eeq
With an abuse of notation (not distinguishing vectors and 1-forms) the dual frame is
\begin{eqnarray}
{\bf e}^1&=&{\bf n}={\bf e}_2\ ,\phantom{--} \qquad {\bf e}^2={\bf l}={\bf e}_1\ , \nonumber \\
{\bf e}^3&=&-{\bf m^*}=-{\bf e}_4 , \qquad {\bf e}^4=-{\bf m}=-{\bf e}_3\ .
\end{eqnarray}
The basis vectors, thought of as directional derivative operators when acting on scalars, are denoted  by
\begin{eqnarray}
{\bf e}_1&=&{\bf e}^2=D , \qquad {\bf e}_2={\bf e}^1=\Delta \ , \nonumber \\
{\bf e}_3&=&-{\bf e}^4=\delta , \qquad {\bf e}_4=-{\bf e}^3=\delta^* ,
\end{eqnarray}
and the 12 complex spin coefficents are introduced through the following linear combinations of the 24  Ricci rotation (connection) coefficents
$\gamma_{cab} = e_c \cdot \nabla_{e_b} e_a = -\gamma_{acb}$
\begin{eqnarray}
\meqalign{& \kappa=\gamma_{311}, &
\,\,\,\,\,\,\,\,\,\,& \rho=\gamma_{314}, &\,\,\,\,\,\,\,\,\,\,& \epsilon=\frac{1}{2} (\gamma_{211}+\gamma_{341}), \cr
& \sigma=\gamma_{313}, &\,\,\,\,& \mu=\gamma_{243}, &\,\,\,\,& \gamma=\frac{1}{2} (\gamma_{212}+\gamma_{342}), \cr
& \lambda=\gamma_{244}, &\,\,\,\,& \tau=\gamma_{312}, &\,\,\,\,& \alpha=\frac{1}{2} (\gamma_{214}+\gamma_{344}), \cr
& \nu=\gamma_{242}, &\,\,\,\,& \pi=\gamma_{241}, &\,\,\,\,& \beta=\frac{1}{2} (\gamma_{213}+\gamma_{343})\ .\,\,\, \cr
}\label{SPINCO}
\end{eqnarray}
As a general rule, complex conjugation of null frame components is equivalent to interchanging the indices 3 and 4.
Using this prescription the inverse relations of (\ref{SPINCO}) are
\begin{eqnarray}
\meqalign{
&\gamma_{121}=-(\epsilon+\epsilon^*),\quad &\gamma_{122}=-(\gamma+\gamma^*),\quad &\gamma_{123}=-(\beta+\alpha^*),\quad &\gamma_{124}=-(\alpha+\beta^*),  \cr
&\gamma_{131}=-\kappa,&\gamma_{132}=-\tau, &\gamma_{133}=-\sigma,&\gamma_{134}=-\rho,   \cr
&\gamma_{141}=-\kappa^*,&\gamma_{142}=-\tau^*, &\gamma_{143}=-\rho^*,&\gamma_{144}=-\sigma^*,   \cr
&\gamma_{231}=\pi^*,&\gamma_{232}=\nu^*,&\gamma_{233}=\lambda^* ,&\gamma_{234}=\mu^*,   \cr
&\gamma_{241}=\pi,&\gamma_{242}=\nu, &\gamma_{243}=\mu,&\gamma_{244}=\lambda,   \cr
&\gamma_{341}=(\epsilon-\epsilon^*),&\gamma_{342}=\gamma-\gamma^*, &\gamma_{343}=\beta-\alpha^*,
&\gamma_{344}=\alpha-\beta^* \,. \cr
}\label{SPINRICCI}
\end{eqnarray}

The commutation rules (Lie brackets) $[e_a,e_b]=C^c{}_{ab} e_c$ reduce to
\beq
\label{comrel}
\meqalign{
[\Delta, D]&=& (\gamma+\gamma^*)D+(\epsilon+\epsilon^*)\Delta-(\tau^*+\pi)\delta -(\tau+\pi^*)\delta^*, \cr
[\delta, D]&=&(\alpha^*+\beta-\pi^*)D+\kappa \Delta -(\rho^*+\epsilon -\epsilon^*)\delta-\sigma \delta^*, \cr
[\delta, \Delta]&=&-\nu^* D+(\tau-\alpha^*-\beta)\Delta +(\mu-\gamma+\gamma^*) \delta +\lambda^* \delta^*, \cr
[\delta^*, \delta]&=& (\mu^*-\mu)D+(\rho^*-\rho)\Delta+(\alpha-\beta^*)\delta +(\beta-\alpha^*)\delta^* . \cr
}
\eeq

The 10 independent components $C_{abcd}$ of the Weyl tensor
are represented by 5 complex scalars in the Newman-Penrose formalism 
\beq
\meqalign{
\Psi_0 &=&-C_{1313}&=& - C_{abcd}~ l^a m^b l^c m^d, \cr
\Psi_1 &=& -C_{1213}&=&- C_{abcd}~l^a n^b l^c m^d, \cr\
\Psi_2 &=&-C_{1342}&=& - C_{abcd}~ l^a  m^b  m^{*c} n^d, \cr
\Psi_3 &=& -C_{1242}&=&- C_{abcd}~l^a n^b m^{*c} n^d, \cr
\Psi_4 &=& -C_{2424}&=&- C_{abcd}~n^a m^{*b} n^c m^{*d}\ ,\cr
}\label{WEYLNP}
\eeq 
with the additional properties
\beq
\meqalign{
C_{1334}&=&C_{1231}=\Psi_1,             
&\quad & C_{1241}&=&C_{1443}=\Psi_1^* ,  \cr
C_{1212}&=&C_{3434}=-(\Psi_2+\Psi_2^*), 
&\quad & C_{1234}&=&(\Psi_2-\Psi_2^*),   \cr
C_{2443}&=&-C_{1242}=\Psi_3,            
&\quad & C_{1232}&=&C_{2343}=-\Psi_3^*,  \cr
\label{WEYLINV}
}
\eeq
and $C_{1314}=C_{2324}=C_{1332}=C_{1442}=0$.
Analogously, the 10 independent components $R_{ab}$ of the Ricci tensor 
are represented by the following (four real and three independent complex) scalars, packaged in their tracefree ($\Phi$) and pure trace ($\Lambda$) parts
\beq\meqalign{
\Phi_{00}&=& -\frac12 R_{11}, &\quad &  
\Phi_{22}&=& -\frac12 R_{22},\cr        
\Phi_{02}&=& -\frac12 R_{33}, &\quad &  
\Phi_{20}&=& -\frac12 R_{44},\cr        
\Phi_{11}&=& -\frac14 ( R_{12}+R_{34} ), &\quad &  
\Phi_{01}&=& -\frac12 R_{13},\cr        
\Phi_{10}&=& -\frac12 R_{14}, &\quad &  
\Phi_{12}&=& -\frac12 R_{23}, \cr       
\Phi_{21}&=& -\frac12 R_{24}, &\quad &  
\Lambda&=&\frac{1}{24}R= -\frac{1}{12}( R_{12}-R_{34} )\ . 
\cr
}\eeq
The components $R_{abcd}$ of the Riemann tensor are then related to the Weyl and Ricci tensors by:
\beq\meqalign{
R_{1212}&=&-\Psi_2-\Psi_2^*-2\Phi_{11}-10 \Lambda , 
&\quad &  R_{1324}&=& \Psi_2+2\Lambda ,\cr
R_{1234}&=& \Psi_2-\Psi_2^*, 
&\quad &  R_{3434}&=& -\Psi_2-\Psi_2^*+2\Phi_{11}-10 \Lambda ,\cr
R_{1313}&=&-\Psi_0 , 
&\quad &  R_{2323}&=& -\Psi_4^*,\cr
R_{1314}&=& -\Phi_{00},
 &\quad &  R_{2324}&=& -\Phi_{22},\cr
R_{3132}&=& \Phi_{02}, 
&\quad &  R_{1213}&=&-\Psi_1-\Phi_{01} ,\cr
R_{1334}&=& -\Psi_1^*-\Phi_{01}, 
&\quad &  R_{1223}&=& \Psi_3^*+\Phi_{12} ,\cr
R_{2334}&=& -\Psi_3^*-\Phi_{12}\, , 
&\quad &  &\quad& \cr
}\eeq
with the additional complex conjugate relations obtained by interchanging the indices 3 and 4.

The six independent components $F_{ab}$ of the (antisymmetric) Maxwell tensor are represented by the three complex scalars
\beq
\phi_0=F_{13}\ ,
\qquad \phi_1=\frac12 (F_{12}+F_{43})\ ,
\qquad \phi_2=F_{42}\ ,
\eeq
with the inverse relations
\begin{eqnarray}
F_{ab}=
\pmatrix{0& \phi_1^* +\phi_1 & \phi_0 & \phi_0^* \cr 
\noalign{\smallskip}
           -\phi_1^* -\phi_1 & 0 & -\phi_2^* & -\phi_2 \cr
\noalign{\smallskip}
           -\phi_0 & \phi_2^* & 0 &\phi_1^*-\phi_1 \cr
\noalign{\smallskip}
           -\phi_0^* & \phi_2 & -\phi_1^*+\phi_1 & 0\cr}\,\,\ .
\end{eqnarray}

Finally the Ricci and the Bianchi identities are given explicitly in the book of Chandrasekhar \cite{chandra} (section 1.8).
Here they will be referred to according to the components of the Riemann tensor which give rise to the corresponding equation:
$[R_{abcd}]$ stands for the Ricci identity associated with $R_{abcd}$ 
and $R_{ab[cd|e]}=0$ for the Bianchi identities.

Exchange of ${\bf l}\leftrightarrow {\bf n}$, ${\bf m}\leftrightarrow {\bf m^{*}}$ in the various Newman-Penrose quantities implies
\begin{eqnarray}
\meqalign{
&D=l^\mu\partial_\mu 
\stackrel{ {\bf l}\leftrightarrow {\bf n}} {\longleftrightarrow}\,\,&
n^\mu\partial_\mu
\stackrel{{\bf m}\leftrightarrow {\bf m^*}}{\longleftrightarrow} 
n^\mu\partial_\mu=\Delta ,\cr
&\delta=m^\mu\partial_\mu 
\stackrel{ {\bf l}\leftrightarrow {\bf n}} {\longleftrightarrow}\,\,&
m^\mu\partial_\mu
\stackrel{{\bf m}\leftrightarrow {\bf m^*}}{\longleftrightarrow} 
m^{*\mu}\partial_\mu=\delta^* ,\cr
}\label{SCAM0}
\end{eqnarray}
\begin{eqnarray}
\meqalign{
&\kappa=\gamma_{311}
\stackrel{ {\bf l}\leftrightarrow {\bf n}} {\longleftrightarrow}\,\,
&\gamma_{322}
\stackrel{{\bf m}\leftrightarrow {\bf m^*}}{\longleftrightarrow}
\gamma_{422}=-\gamma_{242}=-\nu ,\cr
&\tau=\gamma_{312}
\stackrel{ {\bf l}\leftrightarrow {\bf n}} {\longleftrightarrow}
&\gamma_{321}
\stackrel{{\bf m}\leftrightarrow {\bf m^*}}{\longleftrightarrow}
\gamma_{421}=-\gamma_{241}=-\pi ,\cr
&\sigma=\gamma_{313}
\stackrel{ {\bf l}\leftrightarrow {\bf n}} {\longleftrightarrow}
&\gamma_{323}
\stackrel{{\bf m}\leftrightarrow {\bf m^*}}{\longleftrightarrow}
\gamma_{424}=-\gamma_{244}=-\lambda ,\cr
&\rho=\gamma_{314}
\stackrel{ {\bf l}\leftrightarrow {\bf n}} {\longleftrightarrow}
&\gamma_{324}
\stackrel{{\bf m}\leftrightarrow {\bf m^*}}{\longleftrightarrow}
\gamma_{423}=-\gamma_{243}=-\mu ,\cr
}\label{SCAM1}
\end{eqnarray}
\begin{eqnarray}
\meqalign{
&\epsilon=\frac{1}{2}(\gamma_{211}+\gamma_{341})
\stackrel{ {\bf l}\leftrightarrow {\bf n}} {\longleftrightarrow}
&\frac{1}{2}(\gamma_{122}+\gamma_{342})
\stackrel{{\bf m}\leftrightarrow {\bf m^*}}{\longleftrightarrow}
\frac{1}{2}(\gamma_{122}+\gamma_{432})=-\gamma ,\cr
&\alpha=\frac{1}{2}(\gamma_{214}+\gamma_{344})
\stackrel{ {\bf l}\leftrightarrow {\bf n}} {\longleftrightarrow}
&\frac{1}{2}(\gamma_{124}+\gamma_{344})
\stackrel{{\bf m}\leftrightarrow {\bf m^*}}{\longleftrightarrow}
\frac{1}{2}(\gamma_{123}+\gamma_{433})=-\beta ,\cr
}\label{SCAM2}
\end{eqnarray}
\begin{eqnarray}
\meqalign{
&\Psi_0=-C_{1313}
\stackrel{ {\bf l}\leftrightarrow {\bf n}} {\longleftrightarrow}\,\,
&-C_{2323}
\stackrel{{\bf m}\leftrightarrow {\bf m^*}}{\longleftrightarrow}
-C_{2424}=\Psi_4 ,\cr
&\Psi_1=-C_{1213}
\stackrel{ {\bf l}\leftrightarrow {\bf n}} {\longleftrightarrow}\,\,
&-C_{2123}
\stackrel{{\bf m}\leftrightarrow {\bf m^*}}{\longleftrightarrow}
-C_{2124}=-C_{1242}=\Psi_3 ,\cr
&\Psi_2=-C_{1342}
\stackrel{ {\bf l}\leftrightarrow {\bf n}} {\longleftrightarrow}\,\,
&-C_{2341}
\stackrel{{\bf m}\leftrightarrow {\bf m^*}}{\longleftrightarrow}
-C_{2431}=-C_{1342}=\Psi_2 \,\,\,.\cr
}\label{SCAM3}
\end{eqnarray}

\begin{equation}
\meqalign{
\phi_0=F_{13}\stackrel{ {\bf l}\leftrightarrow {\bf n}} {\longleftrightarrow}\,\,&F_{23}\stackrel{{\bf m}\leftrightarrow {\bf m^*}}{\longleftrightarrow}F_{24}=-\phi_2 ,\cr
\phi_1=\frac{1}{2}(F_{12}+F_{43})\stackrel{ {\bf l}\leftrightarrow {\bf n}} {\longleftrightarrow}\,\,&\frac{1}{2}(F_{21}+F_{43})\stackrel{{\bf m}\leftrightarrow {\bf m^*}}{\longleftrightarrow}\frac{1}{2}(F_{21}+F_{34})=-\phi_1 . 
}\label{SCAMMAX}
\end{equation}

These operations are useful since they may be used to extend a given component calculation to other components.

\section{NP gravitational perturbations }
The study of  perturbations of the various fields in the NP formalism is achieved by splitting all the relevant quantities in the form $l = l^A + l^B +..., \Psi_4 = \Psi _{4}^{A} + \Psi _{4}^{B}+...$,\, $\sigma=\sigma^A+\sigma^B+...$,\,$D=D^A+D^B+...$, etc.,  where the {\it A} terms are the background and  the {\it B}'s are small perturbations etc.
The full set of perturbative equations is obtained inserting first of all these splitted quanties in the basic equations of the theory (Ricci and Bianchi identities, Maxwell, Dirac, Rarita-Schwinger equations etc..) and keeping only first order terms.
After some ``ad hoc" algebraic manipulations, which generate second order differential relations, one usually obtains second order coupled linear PDE for the curvature quantities. However for some very special spacetimes, as for Petrov type D ones (i.e. Kerr metric), some PDE's are decoulpled.
In the following we'll present an example of this procedure, applied to a generic vacuum type D metric, for which it follows that:
\begin{eqnarray}
\Psi_{0}^{A} = \Psi_{1}^{A} = \Psi_{3}^{A}  = \Psi_{4}^{A}  =0 \label{NPprima}
\end{eqnarray}
\begin{eqnarray}
\kappa^{A} = \sigma^{A} = \nu^{A}  = \lambda^{A}  =0\,\,\,. \label{NPseconda}
\end{eqnarray}
We start from the exact NP relations:\cite{Teukolsky}:
\begin{eqnarray}
(\delta^{*}-4\alpha+\pi)\Psi_{0}-(D-4\rho-2\epsilon)\Psi_{1}-3\kappa\Psi_{2}=(\delta+\pi^{*}-2\alpha^{*}-2\beta)\Phi_{00} \nonumber\\ -(D-2\epsilon-2\rho^{*})\Phi_{01} +2\sigma\Phi_{10}-2\kappa\Phi_{11}-\kappa^{*}\Phi_{02} \,\label{grav1} 
\end{eqnarray}
\begin{eqnarray}
(\Delta-4\gamma+\mu)\Psi_{0}-(\delta-4\tau-2\beta)\Psi_{1}-3\sigma\Psi_{2}=(\delta+2\pi^{*}-2\beta)\Phi_{01}\nonumber\\ 
-(D-2\epsilon+2\epsilon^{*}-\rho^*)\Phi_{02}-\lambda^{*}\Phi_{00}+2\sigma\Phi_{11}-2\kappa\Phi_{12}\, \label{grav2}
\end{eqnarray}
\begin{eqnarray}
(D-\rho-\rho^{*}-3\epsilon+\epsilon^{*})\sigma-(\delta-\tau+\pi^{*}-\alpha^{*}-3\beta)\kappa-\Psi_{0}=0\,\,\,. \label{grav3}
\end{eqnarray}
In the tetrad language, (\ref{grav1}) represents $R_{13[13\vert4]}=0$,  (\ref{grav2}) represents $R_{13[13\vert2]}=0$ (they are two components of the Bianchi identities) and  (\ref{grav3}) is the Ricci identity component $R_{1313}=0$.
In this notation we have:
\begin{equation}\Phi_{00} \equiv -\frac{1}{2}R_{\mu\nu}\; l^\mu l^\nu = 4\pi T_{\mu\nu} l^\mu l^\nu \equiv 4\pi T_{ll} \label{Tll} 
\end{equation} 
and in the same sense one has to look at the other $\Phi_{ij}$. 
Obviously, in (\ref{Tll})  $\pi$ is not a spin coefficient but it is the usual mathematical constant coming from the right hand side of the Einstein equations. Because we are in vacuo, after the perturbative expansion, in these relations we have that  $\Psi_{0}^{A}$ , $\Psi_{1}^{A}$ , $\sigma^A$ , $\kappa^A$, $\Lambda$ and all the $\Phi_{nm}^A$ are zero and consequently we get:
\begin{eqnarray}
\meqalign{
&(\delta^{*}-4\alpha+\pi)^A\Psi_{0}^B-(D-4\rho-2\epsilon)^A\Psi_{1}^B-3\kappa^B\Psi_{2}^A=\cr  &4\pi[(\delta+\pi^{*}-2\alpha^{*}-2\beta)^AT_{ll}^B-(D-2\epsilon-2\rho^{*})^AT_{lm}^B] 
}
\label{grav1B}
\end{eqnarray}

\begin{eqnarray}
\meqalign{
&(\Delta-4\gamma+\mu)^A\Psi_{0}^B-(\delta-4\tau-2\beta)^A\Psi_{1}^B-3\sigma^B\Psi_{2}^A=\cr &4\pi[(\delta+2\pi^{*}-2\beta)^AT_{lm}^B -(D-2\epsilon+2\epsilon^{*}-\rho^*)^A T_{mm}^B] 
}
\label{grav2B}
\end{eqnarray}
\begin{eqnarray}
\meqalign{
(D-\rho-\rho^{*}-3\epsilon+\epsilon^{*})^A\sigma^B-(\delta-\tau+\pi^{*}-\alpha^{*}-3\beta)^A\kappa^B-\Psi_{0}^B=0\,\,\,.\cr
}  
\label{grav3B}
\end{eqnarray}
In the following, we'll obmit the $A$ superscript for the background quantities to simplify the notation.
The type D background metric satisfy the relations:
\begin{equation}D\Psi_2=3\rho\Psi_2 \end{equation} \begin{equation}\delta\Psi_2=3\tau\Psi_2\,\,\,,\end{equation}
so multiplying (\ref{grav3B}) by $\Psi_2$ and taking into account that 
\beq
\Psi_2 D \sigma^B=D(\Psi_2 \sigma^B)-\sigma^B D\Psi_2=D(\Psi_2 \sigma^B)-\sigma^B 3\rho\Psi_2
\eeq
\beq
\Psi_2 \delta \kappa^B=\delta(\Psi_2 \kappa^B)-\kappa^B \delta\Psi_2=\delta(\Psi_2 \kappa^B)-\kappa^B 3\tau\Psi_2
\eeq
we obtain the relation:
\begin{eqnarray}
(D-3\epsilon+\epsilon^{*}-4\rho-\rho^{*})\Psi_2\sigma^B-(\delta+\pi^{*}-\alpha^{*}-3\beta-4\tau)\Psi_2\kappa^B-\Psi_{0}^B\Psi_{2}=0 . \,\,\,\,\label{grav4}
\end{eqnarray}
We have now to eliminate $\Psi_1^B$ from the eqns (\ref{grav1B}) and (\ref{grav2B}). 
To accomplish this task we use the commuting relation found by Teukolsky and valid for any type D metric:
\begin{eqnarray}
&&[D-({\bf p}+1)\epsilon+\epsilon^{*}+{\bf q}\rho-\rho^{*}](\delta-{\bf p}\beta+{\bf q}\tau)\nonumber\\
&&-[\delta-({\bf p}+1)\beta-\alpha^{*}+\pi^{*}+{\bf q}\tau](D-{\bf p}\epsilon+{\bf q}\rho)=0 \label{commfond}
\end{eqnarray}
where {\bf p} and {\bf q} are two generic constants. 
We apply now $(D-3\epsilon+\epsilon^{*}-4\rho-\rho^{*})$ on (\ref{grav2B}) and $(\delta+\pi^{*}-\alpha^{*}-3\beta-4\tau)$ on (\ref{grav1B}) and we subtract the two resulting equations. We can now use (\ref{commfond}) with the choice ${\bf p}=2$ and ${\bf q}=-4$ to eliminate the $\Psi_1^B$ term. The combination $\sigma^B$ and $\kappa^B$ remained is exactly the same present in eq. (\ref{grav4}), and consequently these quantities can be eliminate in favor of the $\Psi_2\Psi_0^B$ term.
The final equation is:
\begin{eqnarray}
[(D-3\epsilon+\epsilon^{*}-4\rho-\rho^{*})(\Delta+\mu-4 \gamma)-(\delta+\pi^{*}-\alpha^{*}-3\beta-4\tau)\nonumber \\
\times \, (\delta^{*}+\pi-4\alpha)-3\Psi_{2}]\Psi_0^B = 4\pi T_{0} \,\, \label{disg1}
\end{eqnarray}
where we have posed:
\begin{eqnarray}
T_{0}&=(\delta+\pi^*-\alpha^*-3\beta-4\tau)[(D-2\epsilon-2\rho^*)T_{lm}^{B}-(\delta+\pi^*-2\alpha^*-2\beta)T_{ll}^{B}] \nonumber \\
&+(D-3\epsilon+\epsilon^*-4\rho-\rho^*)[(\delta+2\pi^*-2\beta)T_{lm}^{B}-(D-2\epsilon+2\epsilon^*-\rho^*)T_{mm}^{B}] \,\,\,.\nonumber \\
\end{eqnarray}
In the same way one can build the other relations to perform the perturbative analisys of a type D vacuum spacetime.
We point out that the same technique can be applied to the NP Maxwell equations to decouple them as well to the neutrino and Rarita-Schwinger ones~\cite{Teukolsky,Guven}
It is important to stress that this perturbative theory has a purely ``algebraic and manipulative" nature, without taking into account the geometry underlying the problem. In the following we'll give to this subject a new geometrical meaning, framing it into the context of the field theory.
But to accomplish this task we have to adapt the previous studies to the Kerr solution, which describes a rotating black hole of specific angular momentum $a$ and mass $M$, with the condition of flat spacetime at infinity.

\section{The Teukolsky Master Equation}

The Kerr solution in Boyer-Lindquist coordinates is given by~\cite{MTW}:
\begin{eqnarray}
ds^2 &=  {\left( 1- \frac{2Mr}{\Sigma} \right)} dt^2 +
       \frac{{ 4aMr\sin^2\theta }}{\Sigma} dt d\phi -
       { {\Sigma} \over {\Delta} } dr^2\nonumber \\
&-\Sigma d\theta^2  
- {2Mr\left[r^2+a^2+\frac{a^2\sin^2\theta}{\Sigma}\right]}
       \sin^2\theta d\phi^2
\label{KNMETRIC}
\end{eqnarray}
where as usual:
\begin{eqnarray}\label{DEL}
\Delta \equiv r^2 - 2Mr + a^2, \qquad 
\Sigma \equiv r^2 + a^2\cos^2\theta 
\end{eqnarray} 
For any Petrov type D metric, and in particular for the Kerr solution, simplifications occur working in a tetrad adapted to the two repeated principal null directions of the corresponding Weyl tensor.
In the NP formalism the following quantities \cite{Bose} completely describe the Kerr solution 
(in this section we use the label $A$ over some quantities in view of a perturbative analysis, as will be clear in the following).
The Kinnersley tetrad~\cite{Kinnersley}:
\begin{eqnarray}
&(l^{\mu})^A&=\frac{1}{\Delta}[r^2+a^2,\Delta,0,a]\,\,\,\,\,\,\,\,\,\,\, \nonumber \\
&(n^{\mu})^A&=\frac{1}{2\Sigma}[r^2+a^2,-\Delta,0,a]\,\,\,\,\,\,\,\,\,\, \\
&(m^{\mu})^A&=\frac{1}{\sqrt{2}(r+ia\cos\theta)}\,[{ia}\,{\sin\theta},0,1,\frac{i}{\sin\theta}]\,\,, \nonumber
\label{TEC0}
\end{eqnarray}
with the 4th tetrad vector being $(m^{*\mu})^A$, the complex conjugate of $(m^{\mu})^A$.
The Weyl tensor is represented by:
\begin{eqnarray}\label{PSIKERR}
&\Psi_0^A=\Psi_1^A=\Psi_3^A=\Psi_4^A=0 \nonumber \\
&\,\\
&\Psi_2^A=M(\rho^A)^3\, \nonumber
\end{eqnarray}
while the Ricci tensor and the curvature scalar are identically zero because of the vacuum spacetime.
The spin coefficents are given by:
\begin{eqnarray}
&\,&\kappa^A =\sigma^A=\,\lambda^A=\,\nu^A=\,\epsilon^A=\,0 \,\,,\nonumber \\
&\,&\rho^A =\frac{-1}{(r-ia\cos\theta)},\qquad\qquad 
\tau^A=\frac{-ia\rho^A\rho^{*A}\sin\theta}{\sqrt{2}} \,\,, \nonumber \\
&\,&\beta^A =\frac{-\rho^{*A}\cot\theta}{2\sqrt{2}},\qquad\qquad
\pi^A=\frac{ia(\rho^{A})^2\sin\theta}{\sqrt{2}}\,\,\,, \\
&\,&\mu^A =\frac{(\rho^A)^2\rho^{*A}\Delta}{2},\qquad\qquad
\gamma^A=\mu^A+\frac{\rho^A\rho^{*A}(r-M)}{2}\,\,\,, \nonumber \\
&\,&\alpha^A =\pi^A-\beta^{*A}\,\,\,.\nonumber
\label{TEC2}
\end{eqnarray}
It is a relevant result that all the massless (non trivial) perturbations of different spin of the Kerr metric 
are included into the Teukolsky  Master Equation (TME)~\cite{Teukolsky} which has the following form ($\pi$ is the mathematical constant):
\begin{eqnarray}\label{master2}
&&{}\left [ \frac{(r^2+a^2)^2}{\Delta} - a^2 \sin^2 \theta \right ] \frac{\partial^2 \psi}{\partial t^2} +
\frac{4 M a r}{\Delta} \frac{\partial^2 \psi}{\partial t \partial \phi} + 
\left [ \frac{a^2}{\Delta} - \frac{1}{\sin^2 \theta} \right ] \frac{\partial^2
\psi}{\partial \phi^2} \nonumber \\
&& - \Delta^{-s} \frac{\partial}{\partial r} \left ( \Delta^{s+1}\frac{\partial
\psi}{\partial r}\right ) - \frac{1}{\sin \theta} \frac{\partial}{\partial \theta} \, \left ( \sin \theta
\frac{\partial \psi}{\partial \theta} \right ) -2 s \left [\frac{a (r-M)}{\Delta} + \frac{i \, \cos \theta}{\sin^2
\theta} \right ] \frac{\partial \psi}{\partial \phi}  \\ 
&& - 2 s \left [ \frac{M(r^2 - a^2)}{\Delta} - r
- i\, a\, \cos \theta \right ] \frac{\partial \psi}{\partial t} + \left ( s^2 \cot^2 \theta - s \right ) \psi = 4\pi \Sigma T\,. \nonumber
\end{eqnarray}

The following table shows the various quantities involved in the Teukolsky Master Equation ($T_a$ and $J_a$ are respectively gravitational and electromagnetic sources):
\begin{center}
{\bf TME quantities} \\[1ex]
\end{center}
\beq
\meqalign{
\hline
\psi &\qquad \phantom{++}s &\qquad T \cr \hline 
\hline
\Phi &\qquad \phantom{++}0&\qquad  \hbox{\rm see ref.~\cite{Detweiler}}\cr 
\hline
\chi_0 &\qquad +1/2 &\qquad  \hbox {\rm see ref.~\cite{Wainwright}} \cr 
\hline
\rho^{-1}\chi_1 &\qquad -1/2 &\qquad  \hbox {\rm see ref.~\cite{Wainwright}} \cr   
\hline
\phi_0  &\qquad +1 &\qquad    J_0  \cr 
\hline
\rho^{-2}\phi_2&\qquad -1&\qquad   \rho^{-2} J_{2} \cr 
\hline
\Omega_0  &\qquad +3/2&\qquad  \hbox{\rm see ref.~\cite{Guven}}     \cr 
\hline
\rho^{-3}\Omega_3&\qquad -3/2&\qquad   \hbox{\rm see ref.~\cite{Guven}}  \cr 
\hline
\Psi_0 &\qquad +2 &\qquad  2T_0 \cr 
\hline
\rho^{-4}\Psi_4 &\qquad -2 &\qquad  2\rho^{-4}T_4 \cr 
\hline
}
\eeq
\begin{center}
\small{{Table 1: The physical field component $\psi$, spin weight $s$ and source terms $T$ 
for eq. (\ref{master2})}} 
\end{center} 

It is possible to prove that the other NP quantities not described by this equation are related to trivial perturbations in the Kerr background~\cite{Wald,FackerellIpser}, and consequently the study of the Teukolsky Master Equation alone allows  practically the recovery of all the physical quantities.
The reader can easily verify that eq. (\ref{disg1}) coincides with (\ref{master2}) for $s=2$.
It is remarkable the separability of this equation in the form:
\beq
\psi(t,r,\theta,\phi)=e^{-i\omega t}e^{i m \phi} R(r)Y(\theta),
\eeq
by using the so called ``spin weighted spheroidal harmonics" which, as it is well known from molecular physics~\cite{Leaver}, satisfy the following spectral problem:
\beq ({\it H}_0+{\it H}_1)Y(\theta)=-EY(\theta) \label{MeccQuant}\eeq
with
\begin{eqnarray}
\label{MQSFER}
{\it H}_0=\bigg[{1\over \sin\theta}{d\over d\theta}\bigg(\sin\theta {d\over d\theta}\bigg)-\bigg( {m^2+s^2+2ms\cos\theta\over\sin^2\theta}\bigg)\bigg],
\end{eqnarray}
\begin{eqnarray}     
{\it H}_1=a^2\omega^2\cos^2\theta-2a\omega s\cos\theta, 
\end{eqnarray}
and $E=E_{s,l,m}$ and $Y=Y_{s,l,m}$ are the angular eigenvalues and eigenfunctions which in general can only be 
obtained perturbatively.
At this point, the perturbation of the Kerr black hole has become a typical Quantum Mechanics work in which one has to solve a Sturm-Liouville problem for two coupled radial and angular ODE's.
The Teukolsky Master Equation can be cast in a more compact form by introducing a ``connection vector" whose components are:
\begin{eqnarray}
&\Gamma^t =& -\frac{1}{\Sigma}\left[\frac{M(r^2-a^2)}{\Delta }-(r+ia\cos\theta)\right]\nonumber\\
&\Gamma^r =& -\frac{1}{\Sigma}\,(r-M)\nonumber \\
&\Gamma^\theta =&\,0 \nonumber\\
&\Gamma^\phi =&-\frac{1}{\Sigma }\left[\frac{a(r-M)}{\Delta}+i \frac{\cos\theta}{\sin^2\theta}\right]\, .
\label{eq:SPINNOL}
\end{eqnarray}
It's easy to prove that: 
\begin{equation}
\nabla^\mu \Gamma_\mu = -\frac{1}{\Sigma}\quad\quad,\quad\quad \Gamma^\mu \Gamma_\mu = \frac{1}{\Sigma}\cot^2 \theta +4\Psi_2^A \label{eq:GAMMAPROP} 
\end{equation}
and consequently the Teukolsky Master Equation assumes the form:
\begin{equation}
[(\nabla^\mu+s\Gamma^\mu)(\nabla_\mu+s\Gamma_\mu)-4s^2\Psi_2^A]\psi^{(s)}=4\pi T\, 
\label{eq:bellak}
\end{equation}
where $\Psi_2^A$ is the only non vanishing NP component of the Weyl tensor in the Kerr 
background in the Kinnersley tetrad (\ref{PSIKERR}).
Equation (\ref{eq:bellak}) gives a common structure for these massless fields in the Kerr background variating the "s" index. In fact, the first part in the lhs represents (formally) a D'Alembertian, corrected by taking into account the spin-weight and the second one is a curvature (Weyl) term linked to the ``s" index too. 
This particular form of the Teukolsky Master Equation forces us to extend this analysis in the next sections. 

\section{Generalized wave equations}

The compact form of Teukolsky Master Equation we have shown suggests a connection between the perturbation theory and a sort of generalized wave equations which differ from the standard ones by curvature terms. 
In fact generalized wave operators are know as De Rham-Lichnerowicz Laplacians and the curvature terms which make them different from the ordinary ones are given by the Weitzenb\"ock formulas.
Mostly known examples in electromagnetism are 
\begin{itemize}
\item the wave equation for the vector potential $A_{\mu}$\cite{MTW}:
\begin{eqnarray}
\nabla_\alpha\nabla^\alpha A_{\mu}-R_\mu{}^{\lambda}A_\lambda=-4\pi J_\mu\qquad,\qquad\nabla^\alpha A_\alpha=0\label{AMU}
\end{eqnarray}
\item the wave equation for the Maxwell tensor~\cite{Prasanna}:
\beq
\nabla^\mu\nabla_\mu F_{\nu\lambda}+R_{\rho\mu\nu\lambda}F^{\rho\mu}-R^\rho{}_\lambda F_{\nu\rho}+R^\rho{}_\nu F_{\lambda\rho}=-8\pi\nabla_{[\mu }J_{\nu]}
\label{FMUNU}
\eeq
\end{itemize}
while for the gravitational case one has
\begin{itemize}
\item the wave equation for the metric perturbations (gravitational waves)~\cite{MTW}:
$$
\nabla_\alpha\nabla^\alpha \bar{h}_{\mu\nu}+2R_{\alpha\beta\mu\nu}\bar{h}^{\alpha\beta}-2R_{\alpha(\mu}\bar{h}_{\nu)}{}^\alpha=0,\quad
$$
\beq
\quad \nabla_\alpha \bar{h}_\mu{}^\alpha=0,\quad\quad \bar{h}_{\mu\nu}=h_{\mu\nu}-\frac{1}{2}g_{\mu\nu}h_\alpha{}^\alpha \label{HMUNU}
\eeq
\item the wave equation for the Riemann Tensor~\cite{MTW} (in vacuum for simplicity)
\beq
\nabla_\mu \nabla^\mu R_{\alpha\beta\gamma\delta}-R_{\alpha\beta\rho\sigma}R_{\gamma\delta}{}^{\rho\sigma}-2(R_{\alpha\rho\gamma\sigma}R_{\beta}{}^\rho{}_\delta{}^\sigma-R_{\alpha\rho\delta\sigma}R_{\beta}{}^\rho{}_\gamma{}^\sigma)=0\,.\label{RMN}
\eeq
\end{itemize}
These equations are ``non minimal," in the sense that they cannot be recovered by a minimal substitution from their flat space counterparts. A similar situation holds in the standard Quantum Field Theory for the electromagnetic Dirac equation. In fact, applying for instance to the Dirac equation an ``ad hoc" first order differential operator one gets the second order Dirac equation~\cite{Zuber,LL}
\beq
(i \partialslash -e \Aslash+ m)(i \partialslash -e \Aslash- m)\psi=\left[(i \partial_\mu-eA_\mu)(i\partial^\mu-eA^\mu)-\frac{e}{2}\sigma^{\mu\nu}F_{\mu\nu}-m^2\right]\psi=0,
\label{emdirac}
\eeq
where the notation is obvious.
It is easy to recognize in eq. (\ref{emdirac}) a generalized Laplacian and a curvature (Maxwell) term 
applied to the spinor. 
Moreover this equation is ``non minimal",  in the sense that the curvature (Maxwell) term cannot be recoverered by electromagnetic minimal substitution in the standard Klein-Gordon equation for the spinor components. 
The analogous second order Dirac equation in presence of a gravitational field also has a non minimal curvature term~\cite{Pagels,Birrell} and reduces to the form:
\beq ( \nabla_{\alpha} \nabla^{\alpha} + m^2+\frac{1}{4} R ) \psi = 0 \eeq 
Quite recently it was shown by Mohanty and Prasanna~\cite{Prasanna}, that the Maxwell tensor wave equation (\ref{FMUNU}) could be  framed into the context of the QED vacuum polarization in a background gravitational field.
They start from a classical work of Drummond and Hathrell~\cite{Drummond} which
derives higher derivative coupling arising from the QED loop corrections to the graviton-photon vertex, encoded in the Lagrangian, valid for $\omega^2<\alpha/m_e^2$ ($\alpha$ is the fine structure constant and $m_e$ is the electron mass):  

\begin{equation}
{\cal L} =- \frac{1}{4} F_{\mu\nu}F^{\mu\nu} + a R F_{\mu\nu} F^{\mu\nu} +
b R_{\mu\nu} F^{\mu \sigma} F^{\nu}{}_\sigma + c R_{\mu\nu\sigma\tau} F^{\mu\nu} F^{\sigma \tau}
\label{modlag}
\end{equation}
with the coefficients $a = - \frac{5}{720} \frac{\alpha}{\pi}$, $ b
= \frac{26}{720} \frac{\alpha}{\pi }$, $c = - \frac{2}{720}
\frac{\alpha}{\pi }$. 
The physics of these ``tidal" terms is that the effect of virtual electron loops gives to the photon a typical size proportional to the Compton wavelenght of the electron $\lambda_c$. 
The field equations obtained from the modified lagrangian 
(\ref{modlag}) are:
\be
\nabla_\mu F^{\mu\nu} + \Omega^\nu = 0
\ee
where $\Omega^\nu$ represents the non linear quantum correction the explicit (long) form of which is given in~\cite{Prasanna}. 
Manipulating the previous relation one gets the wave equation: 
\beq
\nabla^\mu\nabla_\mu F_{\nu \lambda} + \left\{ R_{\rho \mu \nu \lambda}
F^{\rho\mu} + R^\rho_{\;\;\lambda} F_{\nu\rho} - R^\rho_{\;\;\nu} F_{\lambda \rho} \right\}
+\left[ \nabla_\nu \Omega_\lambda - \nabla_\lambda \Omega_\nu \right] = 0\ .
\label{MOHAP}
\eeq
In the range of validity of the effective action the Riemann and Ricci terms in the
curl brackets of (\ref{MOHAP}) which are present in Einstein's gravity are larger than the terms in the square brackets, arising from the loop corrections, and consequently it is clear that the classical second order wave equation for the Maxwell tensor describes with good accuracy this quantum phenomenon too.

\section{Vacuum gravitational fields}
The Weyl and the Riemann tensor are connected by the relation:
\begin{eqnarray}
R_{\alpha\beta\gamma\delta}&=&C_{\alpha\beta\gamma\delta}+\frac{1}{2} ( g_{\alpha \gamma} R_{\beta \delta}- g_{\beta \gamma} R_{\alpha \delta}- g_{\alpha \delta}R_{\beta \gamma}+ g_{\beta \delta}R_{\alpha \gamma})\nonumber \\ 
&&-\frac{1}{6}( g_{\alpha \gamma}g_{\beta \delta}-g_{\alpha \delta}g_{\beta \gamma})R \ ,
\end{eqnarray}
In a vacuum spacetime, $R_{\alpha\beta}=0, R=0$ and consequently the Riemann tensor coincides with the Weyl one.
The Weyl tensor in vacuum satisfies the generalized de Rham wave equation (the vertical bar indicates the covariant derivative)  
\begin{equation}
\meqalign{
&\Delta_{\rm (dR)} C_{a_1a_2a_3a_4}
= C_{a_1a_2a_3a_4}\, {}_{\vert b}{}^{\vert b}
+(C^{bc}{}_{a_1a_2}C_{bca_3a_4}
+ 4C^b{}_{a_1}{}^c {}_{[a_4}C_{a_3]cba_2}) 
=0\ .\cr
\label{gggi}
}
\end{equation}
The name  ``generalized de Rham wave equation" comes from the nice compact form which can be given to this equation
\beq
\Delta_{\rm (dR)}C_{a_1a_2a_3a_4}
=(\delta D+D \delta ) C_{a_1a_2a_3a_4} =0 \ ,
\eeq
by using the divergence operator $\delta$ and the covariant exterior derivative $D$ \cite{Schouten,Babourova,Plebanski}
(the divergence operator $\delta$ and the covariant exterior derivative $D$ must not be confused with the NP frame derivatives denoted by the same symbols). 
This is the most natural generalization to tensor-valued forms 
of the ordinary de Rham wave operator
\beq
\Delta_{\rm (dR)}=\delta d +d \delta ,
\eeq
whose action is defined only on ordinary differential forms.
A detailed study of the generalized de Rham operator will be discussed elsewhere \cite{bcjr}.

The explicit form taken by $[\Delta_{\rm (dR)}C]_{abcd}=0$ in the Newman-Penrose formalism is now derived and simplified somewhat using the commutator identities to eliminate some second derivatives and the Bianchi and Ricci identities to eliminate some coupling terms between different Weyl tensor scalars.

$\bullet\qquad$ \fbox{$[\Delta_{\rm (dR)}C]_{1313}=0 $} \qquad {\it (wave equation for $\Psi_0$)} 

Recalling that $\Psi_0 =-C_{1313}$, select the ${1313}$ component of eq.~(\ref{gggi}) 
\begin{eqnarray}
\meqalign{
&[-D\Delta-\Delta D+\delta^* \delta+\delta\delta^*
+(9\gamma+\gamma^*-\mu-\mu^*)D+(7\epsilon-\epsilon^*+\rho^*+\rho)\Delta\cr
&+(\beta^*+\pi-9\alpha-\tau^*)\delta+(\pi^*-\tau-7\beta-\alpha^*)\delta^*
+4(D\gamma+\Delta\epsilon-\delta\alpha-\delta^*\beta)\cr
&+4(\alpha\alpha^*-\rho^*\gamma+\tau^*\beta+\tau\pi-\beta^*\beta
-8\gamma\epsilon-\epsilon\gamma^*+\gamma\epsilon^*
-\alpha\pi^*+\mu\epsilon-\beta\pi-\sigma\lambda\cr
&+\kappa\nu+\tau\alpha+8\alpha\beta-\rho\mu
-\rho\gamma+\mu^*\epsilon+3\Psi_2)]\Psi_0
+4[2\rho\delta+2\sigma\delta^*-2\tau D-2\kappa\Delta\cr
&+\delta\rho-D\tau+\delta^*\sigma-\Delta\kappa+\pi^*\rho
-\rho\alpha^*+5\tau\epsilon-\tau\epsilon^*-5\rho\beta 
+\tau\rho^*+\sigma\pi-\kappa\mu^*\cr
&-\kappa\mu-7\sigma\alpha+\kappa\gamma^*+7\kappa\gamma
-\sigma\tau^*+\sigma\beta^*-3\Psi_1]\Psi_1
+[24(\sigma\rho-\kappa\tau)]\Psi_2=0 \ .}\label{DR1313}
\end{eqnarray}
This equation only contains second order directional derivatives of $\Psi_0$, which can be rewritten first using the identities
\begin{eqnarray}
\label{eq:dirder}
D\Delta+\Delta D &=& 2 D\Delta + [\Delta,D]\nonumber \\
\delta\delta^*+\delta^*\delta&=& 2\delta\delta^*+[\delta^*,\delta]
\end{eqnarray}
and then the commutators in these expressions can be replaced by first order derivatives using the commutator identities (\ref{comrel})$_1$ and (\ref{comrel})$_4$, in order to reduce the number of second order differential operators acting on $\Psi_0$.

Equation (\ref{DR1313}) then becomes
\begin{eqnarray}
\meqalign{
&[-2D\Delta+2\delta\delta^*+(8\gamma-2\mu)D+(6\epsilon-2\epsilon^*+2\rho^*)\Delta+(2\pi-8\alpha)\delta\cr
&+(2\pi^*-6\beta-2\alpha^*)\delta^*+4(D\gamma+\Delta\epsilon-\delta\alpha-\delta^*\beta)+4(\alpha\alpha^*-\rho^*\gamma\cr
&+\tau^*\beta+\tau\pi-\beta^*\beta-8\gamma\epsilon-\epsilon\gamma^*+\gamma\epsilon^*-\alpha\pi^*+\mu\epsilon-\beta\pi-\sigma\lambda\cr
&+\kappa\nu+\tau\alpha+8\alpha\beta-\rho\mu-\rho\gamma+\mu^*\epsilon+3\Psi_2)]\Psi_0+4[2\rho\delta+2\sigma\delta^*-2\tau D-2\kappa\Delta\cr
&+\delta\rho-D\tau+\delta^*\sigma-\Delta\kappa+\pi^*\rho-\rho\alpha^*+5\tau\epsilon-\tau\epsilon^*-5\rho\beta +\tau\rho^*+\sigma\pi-\kappa\mu^*\cr
&-\kappa\mu-7\sigma\alpha+\kappa\gamma^*+7\kappa\gamma-\sigma\tau^*+\sigma\beta^*-3\Psi_1]\Psi_1+[24(\sigma\rho-\kappa\tau)]\Psi_2=0\ }\label{DRBIS1313}
\end{eqnarray}

Next the Bianchi identities can be used
to simplify this expression and introduce terms involving $\Psi_0$ to decouple it whenever possible.
This task is achieved by the elimination of two directional derivatives of $\Psi_1$ in 
eq.~(\ref{DRBIS1313}) using two Bianchi identities
\begin{eqnarray}
R_{13[13|4]}=0:\qquad
&&D\Psi_1-\delta^*\Psi_0+(4\alpha-\pi)\Psi_0
-2(2\rho+\epsilon)\Psi_1+3\kappa\Psi_2=0\ , \nonumber\\
R_{13[13|2]}=0:\qquad
&&\delta\Psi_1-\Delta\Psi_0+(4\gamma-\mu)\Psi_0
-2(2\tau+\beta)\Psi_1+3\sigma\Psi_2=0\ .
\end{eqnarray}
Equation (\ref{DRBIS1313}) then becomes
\begin{eqnarray}
\meqalign{
&[-2D\Delta+2\delta\delta^*+2(4\gamma-\mu)D
+2(3\epsilon-\epsilon^*+\rho^*+4\rho)\Delta+2(\pi-4\alpha)\delta\cr
&+2(\pi^*-3\beta-\alpha^*-4\tau)\delta^*+4(D\gamma+\Delta\epsilon
-\delta\alpha-\delta^*\beta)+4(\alpha\alpha^*-\rho^*\gamma\cr
&+\tau^*\beta-\tau\pi-\beta^*\beta-8\gamma\epsilon
-\epsilon\gamma^*+\gamma\epsilon^*-\alpha\pi^*+\mu\epsilon-\beta\pi
-\sigma\lambda\cr
&+\kappa\nu+9\tau\alpha+8\alpha\beta+\rho\mu
-9\rho\gamma+\mu^*\epsilon+3\Psi_2)]\Psi_0
+4[2\sigma\delta^*-2\kappa\Delta\cr
&+\delta\rho-D\tau+\delta^*\sigma-\Delta\kappa+\pi^*\rho
-\rho\alpha^*+\tau\epsilon-\tau\epsilon^*-\rho\beta 
+\tau\rho^*+\sigma\pi-\kappa\mu^*\cr
&-\kappa\mu-7\sigma\alpha+\kappa\gamma^*+7\kappa\gamma
-\sigma\tau^*+\sigma\beta^*-3\Psi_1]\Psi_1=0\ .}
\label{DRTRIS1313}
\end{eqnarray}

Finally this (exact) equation can be given a Teukolsky-like form which is very helpful in type D geometries where most of the known results have been derived. To accomplish this additional step,
multiply the equation by $-1/2$ and use three Ricci identities in two different ways:
a) in (\ref{DRTRIS1313}) first use the identities
\begin{eqnarray}\meqalign{
[1/2 (R_{1212}-R_{3412})]:\qquad
&\Delta\epsilon=D\gamma-\alpha(\tau+\pi^*)-\beta(\tau^*+\pi)+\cr
&+\gamma(\epsilon+\epsilon^*)+\epsilon(\gamma+\gamma^*)-\tau\pi+\nu\kappa-\Psi_2\ ,\cr
[1/2 (R_{1234}-R_{3434})] :\qquad
&\delta^*\beta=\delta\alpha-(\mu\rho-\lambda\sigma)-\alpha\alpha^*-\beta\beta^*+ \cr
&+ 2\alpha\beta-\gamma(\rho-\rho^*)-\epsilon(\mu-\mu^*)+\Psi_2\ ;\cr
}\label{RICOK1}\end{eqnarray}
b) rewrite the other Ricci identity $[R_{2431}]$ in the following form
\beq
[R_{2431}]:\qquad
D\mu-\delta\pi-(\rho^*\mu+\sigma\lambda)-\pi(\pi^*-\alpha^*+\beta)+\mu(\epsilon+\epsilon^*)+\nu\kappa-\Psi_2=0\ , 
\label{RICOK2}
\eeq
multiply it by $\Psi_0$ and 
c) add this (which is actually zero) to the wave equation.
The result is
\begin{eqnarray}
\meqalign{
&[(D-3\epsilon+\epsilon^{*}-4\rho-\rho^{*})(\Delta+\mu-4 \gamma)
-(\delta+\pi^{*}-\alpha^{*}-3\beta-4\tau)\cr
&(\delta^{*}+\pi-4\alpha)-3\Psi_{2}]\Psi_0
+3(\sigma\lambda-\kappa\nu)\Psi_0
+2[2\kappa\Delta-2\sigma\delta^*-\delta^*\sigma+\Delta\kappa\cr
&-\delta\rho+D\tau-7\kappa\gamma
-\beta^*\sigma+7\sigma\alpha+\tau^*\sigma+\kappa\mu^*+\kappa\mu
-\rho\pi^*+\tau\epsilon^*+\beta\rho\cr
&-\epsilon\tau-\kappa\gamma^*-\pi\sigma-\tau\rho^*
+\rho\alpha^*+3\Psi_1 ]\Psi_1=0 \ ;
\label{diso1}
}\end{eqnarray}
d) finally use in  eq.~(\ref{diso1}) two other Ricci identities
\begin{eqnarray}\meqalign{
[R_{3143}]: \quad
&\delta\rho-\delta^*\sigma
=\rho(\alpha^*+\beta)-\sigma(3\alpha-\beta^*)+\tau(\rho-\rho^*)
+\kappa(\mu-\mu^*)-\Psi_1\ ,\cr
[R_{1312}]: \quad
&D\tau-\Delta\kappa=\rho(\tau+\pi^*)+\sigma(\tau^*+\pi)
+\tau(\epsilon-\epsilon^*)-\kappa(3\gamma+\gamma^*)+\Psi_1.\cr
\label{RICO13}
}\end{eqnarray} 
The final form for eq.~(\ref{diso1}) is then
\begin{eqnarray}
\meqalign{
&[(D-3\epsilon+\epsilon^{*}-4\rho-\rho^{*})(\Delta+\mu-4 \gamma)
-(\delta+\pi^{*}-\alpha^{*}-3\beta-4\tau)\cr
&(\delta^{*}+\pi-4\alpha)-3\Psi_{2}]\Psi_0+3(\sigma\lambda
-\kappa\nu)\Psi_0+[4\kappa\Delta-4\sigma\delta^*
-4\delta^*\sigma+4\Delta\kappa\cr
&-20\kappa\gamma-4\beta^*\sigma+20\sigma\alpha
+4\tau^*\sigma+4\kappa\mu^*-4\kappa\gamma^*+10\Psi_1 ]\Psi_1=0\ ,
\label{DED}
}\end{eqnarray}
which is the exact form for the same equation for $\Psi_0$ found by Teukolsky in the very special case of a type D geometry in the first order perturbation regime (see below).
Moreover this equation coincides  with that obtained by Stewart and Walker \cite{StewartWalker} following a different approach and in the  Geroch-Held-Penrose (GHP) formalism \cite{GHP}
\beq
\meqalign{
&(\,\Ph ' \Ph - \Ed ' \Ed - \bar\rho'\Ph
-5\rho\Ph'+\bar\tau\Ed+5\tau\Ed'+4\sigma\sigma'
-4\kappa\kappa'-10\Psi_2)\Psi_0\cr
&+[4\Ph'\kappa-4\Ed'\sigma-4(\bar\rho'-2\rho')\kappa
+4(\bar\tau-2\tau')\sigma+10\Psi_1]\Psi_1\cr
&+(-4\sigma \Ph+4\kappa\Ed-12\kappa\tau+12\rho\sigma)\Psi_2=0\ .}
\label{SWL}
\eeq 

The equivalence of (\ref{DED}) and (\ref{SWL}) is now explicitly shown.
Recall the action of the two-index  $\{p,q\}$-operators  $\Ph $, $\Ph'$, $\Ed $, $\Ed'$ on  two-index $\{p,q\}$-objects.
Taking into account the weights of the $\Psi_n$'s (i.e., 
$\{p,q\}=\{4-2n,0\}$), those of the spin coefficients and
the definition of the weighted derivative operator of an object $\eta$ (which increases its $\{p,q\}$ indices)
\begin{eqnarray}
\meqalign{
\Ph  \eta &=(D-p\epsilon-q\epsilon^*)\eta ,  &\qquad {\rm increasing\, of\, type}\, \{+1,+1\}, \cr
\Ph' \eta &=(\Delta-p\gamma-q\gamma^*)\eta , &\qquad {\rm increasing\,of\, type}\, \{-1,-1\}, \cr
\Ed  \eta &=(\delta-p\beta-q\alpha^*)\eta ,  &\qquad {\rm increasing\,of\, type}\,\{+1,-1\}, \cr
\Ed' \eta &=(\delta^*-p\alpha-q\beta^*)\eta , &\qquad  {\rm increasing\,of\, type}\, \{-1,+1\},\cr
}\label{CCFF}\end{eqnarray}
eq.~(\ref{SWL}) becomes:
\beq
\meqalign{
&[(\Delta-5\gamma-\gamma^*)(D-4\epsilon)-(\delta^*-5\alpha+\beta^*)(\delta-4\beta)+\mu^*(D-4\epsilon)\cr
&-5\rho(\Delta-4\gamma)+\tau^*(\delta-4\beta)+5\tau(\delta^*-4\alpha)-4(\sigma\lambda-\kappa\nu)-10\Psi_2]\Psi_0\cr
&+(-4\sigma D+4\kappa\delta-12\kappa\tau+12\rho\sigma)\Psi_2+[4(\Delta-3\gamma-\gamma^*)\kappa\cr
&-4(\delta^*-3\alpha+\beta^*)\sigma-4(-\mu^*+2\mu)\kappa+4(\tau^*+2\pi)\sigma+10\Psi_1)]\Psi_1=0\, ,\cr}
\label{SWL2}
\eeq
where the second order operator has to be rewritten via the commutation rules (\ref{comrel}).

Then, by using the two Ricci identities $1/2[R_{1212}-R_{3412}]$ and $1/2[R_{1234}-R_{3434}]$ (see eq.~(\ref{RICOK1})), adding equation $[R_{2431}]$ 
multiplied by $\Psi_0$ and using the two Bianchi identities $R_{13[43|2]}=0$ and $R_{13[21|4]}=0$, namely:
\begin{eqnarray}
\meqalign{
&\Delta\Psi_1-\delta\Psi_2-\nu\Psi_0-2(\gamma-\mu)\Psi_1+3\tau\Psi_2-2\sigma\Psi_3=0\, , &\qquad R_{13[43|2]}=0\, ,\cr
&\delta^*\Psi_1-D\Psi_2-\lambda\Psi_0+2(\pi-\alpha)\Psi_1+3\rho\Psi_2-2\kappa\Psi_3=0\, , &\qquad R_{13[21|4]}=0\, ,\cr
}\label{MOGG}
\end{eqnarray}
one gets exactly eq.~(\ref{DED}).

$\bullet\qquad$ \fbox{$[\Delta_{\rm (dR)}C]_{2424}=0 $}\hspace{1cm} {\it (wave equation for $\Psi_4$)} 

The equation for $\Psi_4$ can be immediately obtained from the equation for $\Psi_0$ using the exchange symmetry ${\bf l}\leftrightarrow {\bf n}$, ${\bf m}\leftrightarrow {\bf m^{*}}$ of the Newman-Penrose quantities.
Under this operation, eq.~(\ref{DED}) becomes:
\begin{eqnarray}
\meqalign{
&[(\Delta+3\gamma-\gamma^{*}+4\mu+\mu^{*})(D+4\epsilon-\rho)-(\delta^{*}-\tau^{*}+\beta^{*}+3\alpha+4\pi)\cr
&(\delta-\tau+4\beta)-3\Psi_2]\Psi_4+3(\sigma\lambda-\kappa\nu)\Psi_4 +[4\lambda\delta-4\nu D+ 4\delta\lambda-4D\nu\cr
&-20\nu\epsilon-4\alpha^*\lambda+20\lambda\beta+4\pi^*\lambda+4\nu\rho^*-4\nu\epsilon^*+10\Psi_3]\Psi_3=0 \ .\label{disg2}
}\end{eqnarray}

Working in the GHP formalism, 
the corresponding relation found by Stewart and Walker \cite{StewartWalker} is:

\beq
\meqalign{
&[(\,\Ph ' \Ph - \Ed ' \Ed - 
(4\rho'+\bar \rho')\Ph -\rho \Ph' + (4\tau'+\bar \tau )\Ed +\tau \Ed ' +4 \rho\rho' -4\tau\tau' -2 \Psi_2]
\Psi_4 \cr 
&+ [4\Ph \kappa ' -4 \Ed \sigma ' -4 (\bar \rho -2\rho)\kappa' +4 (\bar\tau -2\tau)\sigma' +10 \Psi_3]\Psi_3\cr
& +
[-4\sigma' \Ph' +4 \kappa' \Ed' -12 \kappa' \tau' +12 \rho'\sigma ']\Psi_2=0 .
}
\eeq

$\bullet\qquad$ \fbox{$[\Delta_{\rm (dR)}C]_{1213}=0 $}\hspace{1cm} {\it (wave equation for $\Psi_1$)}

Selecting the component $C_{1213}$ ($\Psi_1 =-C_{1213}$) in eq.~(\ref{gggi}) one has

\begin{eqnarray}\meqalign{
&[\delta\delta^*+\delta^*\delta-D\Delta-\Delta D+(\gamma^*+5\gamma-\mu-\mu^*)D+(\rho+\rho^*+3\epsilon-\epsilon^*)\Delta\cr
&+(\pi+\beta^*-\tau^*-5\alpha)\delta+(\pi^*-3\beta-\tau-\alpha^*)\delta^*+2(D\gamma+\Delta\epsilon-\delta\alpha-\delta^*\beta\cr
&+5\tau\pi+\alpha\alpha^*-4\gamma\epsilon-\rho\gamma-\rho^*\gamma+\mu^*\epsilon-\epsilon\gamma^*-\beta\beta^*+\gamma\epsilon^*-\beta\pi\cr
&-5\sigma\lambda+\beta\tau^*-\alpha\pi^*+\mu\epsilon+4\alpha\beta-5\rho\mu+\tau\alpha+5\kappa\nu-3\Psi_2)]\Psi_1\cr
&+[2\nu D+2\pi\Delta-2\lambda\delta-2\mu\delta^*+D\nu+\Delta\pi-\delta\lambda-\delta^*\mu+\tau^*\mu-\nu\rho^*+\tau\lambda\cr
&+\pi\mu^*+\lambda\alpha^*-7\gamma\pi-5\nu\epsilon+7\alpha\mu-\pi\gamma^*+\nu\epsilon^*-\nu\rho+5\lambda\beta-\lambda\pi^*-\mu\beta^*+6\Psi_3]\Psi_0\cr
&+3[-2\tau D-2\kappa\Delta+2\rho\delta+2\sigma\delta^*-D\tau-\Delta\kappa+\delta\rho+\delta^*\sigma+\beta^*\sigma-\beta\rho-\sigma\tau^*\cr
&-\kappa\mu^*+\kappa\gamma^*-3\sigma\alpha-\tau\epsilon^*+\epsilon\tau-\alpha^*\rho+\tau\rho^*+3\gamma\kappa-\mu\kappa+\sigma\pi+\rho\pi^*]\Psi_2\cr
&+12(\sigma\rho-\kappa\tau)\Psi_3=0\,\,\,. \label{DR1213}
}\end{eqnarray}

Using the commutation rules (\ref{comrel}), multiplying by $-1/2$, and using the six Bianchi identities  $R_{13[13\vert4]}=0$,  $R_{13[13\vert2]}=0$,  $R_{13[21\vert4]}=0$,  $R_{13[43\vert2]}=0$,  $R_{42[13\vert2]}=0$ $R_{42[13\vert4]}=0$ and the five Ricci identities $[R_{3143}]$,  $[R_{1312}]$, $[R_{2443}]$, $[R_{2421}]$, $[\frac{1}{2}(R_{1234}-R_{3434}))]$, eq.~(\ref{DR1213}) becomes:

\begin{eqnarray}
\meqalign{
&[D\Delta-\delta\delta^*+2(\mu-\gamma)D+(-3\rho-\rho^*-\epsilon+\epsilon^*)\Delta+2(\alpha-\pi)\delta+(\beta+\alpha^*+3\tau-\pi^*)\delta^*\cr
&-D\gamma-\Delta\epsilon+2\delta\alpha+2\rho^*\gamma-\gamma\epsilon^*+\alpha\pi^*+\kappa\nu+\epsilon\gamma^*+6\rho\gamma+4\gamma\epsilon-6\rho\mu-7\tau\alpha+5\tau\pi-\beta\tau^*\cr
&-2\alpha\alpha^*-2\alpha\beta+3\beta\pi-4\mu\epsilon+7\Psi_2]\Psi_1+[\lambda\delta-\nu D+\delta\lambda-D\nu-3\tau\lambda+\nu\epsilon-\alpha^*\lambda\cr
&+3\rho\nu-\nu\epsilon^*+\lambda\pi^*+\nu\rho^*-\beta\lambda-2\Psi_3]\Psi_0+3[\kappa\delta-\sigma D+2\sigma\rho+2\kappa\beta-2\kappa\tau-2\sigma\epsilon]\Psi_3\cr
&+3[\Delta\kappa-\delta^*\sigma-2\mu\kappa+2\sigma\pi-\sigma\beta^*+\kappa\mu^*-3\kappa\gamma+\sigma\tau^*-\kappa\gamma^*+3\alpha\sigma]\Psi_2=0 .
\label{DR1213COMPACT}
}
\end{eqnarray}
Now adding to the previous relation the equation $[R_{2431}]$ 
(see eq.~(\ref{RICOK2})) multiplied $2\Psi_1$ and using another Ricci identity, $[\frac{1}{2}(R_{
1212}-R_{3412}))]$, (\ref{DR1213COMPACT}) takes the compact form:

\begin{eqnarray}
\meqalign{
&[(D-\epsilon+\epsilon^*-3\rho-\rho^*)(\Delta+2\mu-2\gamma)-(\delta-\alpha^*-\beta+\pi^*-3\tau)(\delta^*+2\pi-2\alpha)\cr
&+6\Psi_2]\Psi_1+2(\kappa\nu-\sigma\lambda)\Psi_1+[\lambda\delta-\nu D+\delta\lambda-D\nu-3\tau\lambda+\nu\epsilon-\alpha^*\lambda+3\rho\nu\cr
&-\nu\epsilon^*+\lambda\pi^*+\nu\rho^*-\beta\lambda-2\Psi_3]\Psi_0+3[\kappa\delta-\sigma D+2\sigma\rho+2\kappa\beta-2\kappa\tau-2\sigma\epsilon]\Psi_3\cr
&+3[\Delta\kappa-\delta^*\sigma-2\mu\kappa+2\sigma\pi-\sigma\beta^*+\kappa\mu^*-3\kappa\gamma+\sigma\tau^*-\kappa\gamma^*+3\alpha\sigma]\Psi_2=0\ .
\label{DR1213SUPERCOMPACT}
}
\end{eqnarray}

The analogous equation in the GHP formulation is
\beq
\meqalign{
&[(\Ph - \bar \rho -3\rho)(\Ph'-2\rho')-(\Ed-\bar\tau '-3\tau)(\Ed'-2\tau')+2(\sigma\sigma'-\kappa \kappa')+6\Psi_2]\Psi_1\cr
&+[3\kappa \Ed -3\sigma \Ph +6 \sigma\rho -6\kappa \tau]\Psi_3 \cr
&+[\Ph \kappa' + \kappa' \Ph -\sigma' \Ed -\Ed \sigma' +\sigma'\bar \tau ' +3 \sigma' \tau -\kappa'\bar \rho -3 \kappa' \rho -2\Psi_3]\Psi_0 \cr
&+[3\Ph ' \kappa -3 \Ed' \sigma +6 \kappa \rho' -6\sigma \tau' +3 \sigma \bar \tau -3 \kappa \bar \rho ']\Psi_2=0\ .
}
\eeq

$\bullet\qquad$ \fbox{$[\Delta_{\rm (dR)}C]_{1242}=0 $}\hspace{1cm} {(\it wave equation for $\Psi_3$)} 

The equation for $C_{1242}=-\Psi_3$ can be easily obtained, as was done for $\Psi_4$, applying the interchange symmetry ${\bf l}\leftrightarrow {\bf n}$, ${\bf m}\leftrightarrow {\bf m^{*}}$ to (\ref{DR1213SUPERCOMPACT}), yielding:
\begin{eqnarray}\meqalign{
&[(\Delta+\gamma-\gamma^*+3\mu+\mu^*)(D-2\rho+2\epsilon)-(\delta^*+\beta^*+\alpha-\tau^*+3\pi)(\delta-2\tau+2\beta)\cr
&+6\Psi_2]\Psi_3+2(\kappa\nu-\sigma\lambda)\Psi_3+[\kappa\Delta-\sigma\delta^*+\Delta\kappa-\delta^*\sigma-3\sigma\pi+\kappa\gamma-\beta^*\sigma+3\mu\kappa\cr
&-\kappa\gamma^*+\sigma\tau^*+\kappa\mu^*-\alpha\sigma-2\Psi_1]\Psi_4+3[\lambda\Delta-\nu\delta^*+2\lambda\mu+2\nu\alpha-2\nu\pi-2\lambda\gamma]\Psi_1\cr
&+3[\delta\lambda-D \nu-2\rho\nu+2\lambda\tau-\lambda\alpha^*+\nu\rho^*-3\nu\epsilon+\lambda\pi^*-\nu\epsilon^*+3\beta\lambda]\Psi_2=0 \ .
}\label{PSIQUARTO}\end{eqnarray}

The analogous equation in the GHP formulation is

\beq
\meqalign{
&[(\Ph' - \bar \rho'
-3\rho')(\Ph-2\rho)-(\Ed'-\bar\tau-3\tau')(\Ed-2\tau)+2(\sigma'\sigma-\kappa'
\kappa)+6\Psi_2]\Psi_3\cr
&+[3\kappa' \Ed' -3\sigma' \Ph' +6 \sigma'\rho' -6\kappa' \tau']\Psi_1
\cr
&+[\Ph' \kappa + \kappa \Ph' -\sigma \Ed' -\Ed' \sigma +\sigma\bar \tau
+3 \sigma \tau' -\kappa\bar \rho' -3 \kappa \rho' -2\Psi_1]\Psi_4 \cr
&+[3\Ph \kappa' -3 \Ed\sigma' +6 \kappa' \rho -6\sigma' \tau +3 \sigma'
\bar \tau' -3 \kappa' \bar \rho]\Psi_2=0\, .
}
\eeq

$\bullet\qquad$ \fbox{$[\Delta_{\rm (dR)}C]_{1342}=0 $}\hspace{1cm} {(\it wave equation for $\Psi_2$)} 

Selecting the component ${1342}$ of  (\ref{gggi}) ($\Psi_2 =-C_{1342}$) one has
\begin{eqnarray}\meqalign{
&[-D\Delta-\Delta D +\delta\delta^*+\delta^*\delta+(\gamma+\gamma^*-\mu-\mu^*)D+(\rho+\rho^*-\epsilon-\epsilon^*)\Delta\cr
&+(\pi+\beta^*-\alpha-\tau^*)\delta+(\pi^*+\beta-\alpha^*-\tau)\delta^*+12\tau\pi-12\rho\mu+12\nu\kappa-12\lambda\sigma-6\Psi_2]\Psi_2\cr
&+[4\nu D+4\pi\Delta-4\lambda\delta-4\mu\delta^*+2D\nu+2\Delta\pi-2\delta\lambda-2\delta^*\mu+2\mu\tau^*-6\gamma\pi+2\nu\epsilon^*-2\rho^*\nu\cr
&-2\mu\beta^*+2\lambda\beta+2\lambda\alpha^*+2\pi\mu^*-2\pi\gamma^*-2\nu\epsilon+2\tau\lambda-2\rho\nu-2\lambda\pi^*+6\alpha\mu+4\Psi_3]\Psi_1\cr
&+[-4\tau D-4\kappa\Delta+4\rho\delta+4\sigma\delta^*-2D\tau-2\Delta\kappa+2\delta\rho+2\delta^*\sigma+2\pi^*\rho-2\epsilon^*\tau+2\tau\rho^*\cr
&-2\mu\kappa+2\alpha\sigma+6\rho\beta-2\tau^*\sigma+2\sigma\beta^*+2\pi\sigma-2\kappa\mu^*-2\kappa\gamma-6\epsilon\tau-2\rho\alpha^*+2\kappa\gamma^*]\Psi_3\cr
&+[4\rho\sigma-4\tau\kappa]\Psi_4+[4\lambda\mu-4\nu\pi+2\Psi_4]\Psi_0=0 \, .
}\label{DRPSI2}\end{eqnarray}

Using the commutation rules (\ref{comrel}), multiplying (\ref{DRPSI2}) by $-1/2$, and using the four Bianchi identities $R_{13[21\vert4]}=0$, $R_{13[43\vert2]}=0$,  $R_{42[13\vert2]}=0$ $R_{42[13\vert4]}=0$, the four Ricci identities $[R_{3143}]$,  $[R_{1312}]$, $[R_{2443}]$, $[R_{2421}]$ and adding eq.~(\ref{RICOK2}) once multiplied by $3\Psi_2$, relation  (\ref{DRPSI2}) takes the form:
\begin{eqnarray}\meqalign{
&[(D+\epsilon+\epsilon^*-2\rho-\rho^*)(\Delta+3\mu)-(\delta+\pi^*-2\tau+\beta-\alpha^*)(\delta^*+3\pi)]\Psi_2\cr
&+(3\lambda\sigma-3\nu\kappa)\Psi_2+2[-\nu D+\lambda\delta-D\nu+\delta\lambda+\nu\rho^*+\lambda\pi^*-2\tau\lambda\cr
&-\lambda\alpha^*+\lambda\beta-\nu\epsilon^*-\nu\epsilon+2\rho\nu+\Psi_3]\Psi_1+2[\kappa\Delta-\sigma\delta^*+\Delta\kappa\cr
&-\delta^*\sigma+\alpha\sigma+\kappa\mu^*+2\mu\kappa+\tau^*\sigma-\kappa\gamma^*-\gamma\kappa-2\pi\sigma-\sigma\beta^*]\Psi_3-\Psi_0\Psi_4=0\cr
}\label{FINPSI2}\end{eqnarray}
The corresponding equation in the GHP formalism is

\beq
\meqalign{
[(\Ph -2\rho - \bar\rho)(\Ph ' -3\rho ')-(\Ed -2\tau -\bar\tau ')(\Ed ' -3\tau ')+3(\kappa \kappa' -\sigma\sigma')]\Psi_2\cr
+2[\kappa' \Ph +\Ph \kappa' -\sigma' \Ed -\Ed \sigma' +2\tau\sigma' +\sigma' \bar \tau ' -2\rho\kappa ' -\kappa' \bar \rho ]\Psi_1\cr
+2[\kappa \Ph' +\Ph' \kappa -\Ed' \sigma -\sigma \Ed' +\Psi_1 +2\tau'\sigma -2\rho' \kappa +\bar \tau \sigma -\kappa \bar \rho ']\Psi_3-\Psi_0\Psi_4=0\, .
}
\eeq

At this point the complete set of (exact) wave equations satisfied by the Weyl scalars has been given. We can now perform their perturbation expansion. 

\subsection{The perturbation theory}

Introducing the operators:
\begin{eqnarray}\meqalign{
&\widehat{\cal T}_0=[(D-3\epsilon+\epsilon^{*}-4\rho-\rho^{*})(\Delta+\mu-4 \gamma)-(\delta+\pi^{*}-\alpha^{*}-3\beta-4\tau)
(\delta^{*}+\pi-4\alpha)-3\Psi_{2}],\cr
&\widehat{\cal W}_0=3(\sigma\lambda-\kappa\nu),\cr
&\widehat{\cal P}_1=[4\kappa\Delta-4\sigma\delta^*-4\delta^*\sigma+4\Delta\kappa-20\kappa\gamma-4\beta^*\sigma+20\sigma\alpha+4\tau^*\sigma+4\kappa\mu^*-4\kappa\gamma^*+10\Psi_1 ],\cr 
}\end{eqnarray} 
where the letter $\hat{\cal T}_0$ here stands for the operator on $\Psi_0$ found by Teukolsky in the perturbation regime; this is a notation extending that due to Lousto and Campanelli \cite{LoustoCampanelli}. In the same way a generic operator $\hat {\cal \chi}$ acting on $\Psi_i$ will be denoted by $\hat {\cal \chi}_i$.
Thus, eq.~(\ref{DED}) can be written as:
\beq
(\widehat{\cal T}_0+\widehat{\cal W}_0)\Psi_0+\widehat{\cal P}_1\Psi_1=0\, .
\eeq
Performing the standard perturbation expansion of the various quantities {\it \`a la Teukolsky}: 
$\Psi_0=\Psi _{0}^{(0)} + \Psi _{0}^{(1)}+ \Psi _{0}^{(2)}+...$,\, $\sigma=\sigma^{(0)}+\sigma^{(1)}+...$,\,$D=D^{(0)}+D^{(1)}+...$ and so on,
one finds 
\begin{eqnarray}\meqalign{
&[(\widehat{\cal T}_0^{(0)}+\widehat{\cal T}_0^{(1)}+\widehat{\cal T}_0^{(2)}+...)+(\widehat{\cal W}_0^{(0)}+\widehat{\cal W}_0^{(1)}+\widehat{\cal W}^{(2)}+...)](\Psi_0^{(0)}+\Psi_0^{(1)}+\Psi_0^{(2)}+...)\cr
&+(\widehat{\cal P}_1^{(0)}+\widehat{\cal P}_1^{(1)}+\widehat{\cal P}_1^{(2)}+...)(\Psi_1^{(0)}+\Psi_1^{(1)}+\Psi_1^{(2)}+...)=0\, ,
}\end{eqnarray}
giving  an inductive procedure for obtaining  a hierarchy of  
equations of the following type:
\begin{eqnarray}
\meqalign{
&(\widehat{\cal T}_0^{(0)}+\widehat{\cal W}_0^{(0)})\Psi_0^{(0)}+\widehat{\cal P}_1^{(0)}\Psi_1^{(0)}=0,\cr
&(\widehat{\cal T}_0^{(1)}+\widehat{\cal W}_0^{(1)})\Psi_0^{(0)}+(\widehat{\cal T}_0^{(0)}+\widehat{\cal W}_0^{(0)})\Psi_0^{(1)}+(\widehat{\cal P}_1^{(0)}\Psi_1^{(1)}+\widehat{\cal P}_1^{(1)}\Psi_1^{(0)})=0,\cr
&\qquad\vdots\cr
&(\widehat{\cal T}_0^{(n)}+\widehat{\cal W}_0^{(n)})\Psi_0^{(0)}+(\widehat{\cal T}_0^{(0)}+\widehat{\cal W}_0^{(0)})\Psi_0^{(n)}+(\widehat{\cal P}_1^{(n)}\Psi_1^{(0)}+\widehat{\cal P}_1^{(0)}\Psi_1^{(n)})={\cal S}_0^{(n)}\, ,
}\label{DIKM}
\end{eqnarray}
where 
\beq
{\cal S}_0^{(n)}=-\sum_{p=1}^{n-1}[(\widehat{\cal T}_0^{(n-p)}+\widehat{\cal W}_0^{(n-p)})\Psi_0^{(p)}+\widehat{\cal P}_1^{(n-p)}\Psi_1^{(p)}]\, ,\qquad(n \geq 2)\, .
\eeq
For every $n\geq 2$ perturbation equation,  the ``source term" ${\cal S}_0^{(n)}$ is composed of quantities at most of order $(n-1)$, which in principle are known.

For a type D background geometry it is well known that
\begin{eqnarray}\meqalign{
&\kappa^{(0)}=\sigma^{(0)}=\lambda^{(0)}=\nu^{(0)}=0,\cr
&\Psi_0^{(0)}=\Psi_1^{(0)}=\Psi_3^{(0)}=\Psi_4^{(0)}=0,
}\end{eqnarray}
so $\widehat{\cal W}_0^{(0)}=\widehat{\cal W}_0^{(1)}=\widehat{\cal P}_1^{(0)}=0$ and the recurrence scheme (\ref{DIKM}) becomes:
\begin{eqnarray}
\meqalign{
&0=0,\cr
&\widehat{\cal T}_0^{(0)}\Psi_0^{(1)}=0,\cr
&\qquad\vdots\cr
&\widehat{\cal T}_0^{(0)}\Psi_0^{(n)}={\cal S}_0^{(n)}\, .
}
\label{xteuko1}
\end{eqnarray}
One can study the perturbation of eq.~(\ref{disg2}) exactly in the same way as before.
In fact, defining the operators
\begin{eqnarray}\meqalign{
&\widehat{\cal T}_4=[(\Delta+3\gamma-\gamma^{*}+4\mu+\mu^{*})(D+4\epsilon-\rho)-(\delta^{*}-\tau^{*}+\beta^{*}+3\alpha+4\pi)\cr
&\qquad(\delta-\tau+4\beta)-3\Psi_2],\cr
&\widehat{\cal P}_3=[4\lambda\delta-4\nu D+4\delta\lambda-4D\nu-20\nu\epsilon-4\alpha^*\lambda+20\lambda\beta+4\pi^*\lambda+4\nu\rho^*-4\nu\epsilon^*+10\Psi_3],\cr 
&\widehat{\cal W}_4=3(\sigma\lambda-\kappa\nu),
\cr
}\end{eqnarray} (where the letter $\hat{\cal T}_4$ here stands for the operator on $\Psi_4$ found by Teukolsky), eq.~(\ref{disg2})
can be written as
\beq
(\widehat{\cal T}_4+\widehat{\cal W}_4)\Psi_4+\widehat{\cal P}_3\Psi_3=0\,\,.
\eeq
The hierarchy of equations in this case is:
\begin{eqnarray}\meqalign{
&(\widehat{\cal T}_4^{(0)}+\widehat{\cal W}_4^{(0)})\Psi_4^{(0)}+\widehat{\cal P}_3^{(0)}\Psi_3^{(0)}=0, \cr
&(\widehat{\cal T}_4^{(1)}+\widehat{\cal W}_4^{(1)})\Psi_4^{(0)}+(\widehat{\cal T}_4^{(0)}+\widehat{\cal W}_4^{(0)})\Psi_4^{(1)}+(\widehat{\cal P}_3^{(0)}\Psi_3^{(1)}+\widehat{\cal P}_3^{(1)}\Psi_3^{(0)})=0, \cr
&\qquad\vdots\cr
&(\widehat{\cal T}_4^{(n)}+\widehat{\cal W}_4^{(n)})\Psi_4^{(0)}+(\widehat{\cal T}_4^{(0)}+\widehat{\cal W}_4^{(0)})\Psi_4^{(n)}+(\widehat{\cal P}_3^{(n)}\Psi_3^{(0)}+\widehat{\cal P}_3^{(0)}\Psi_3^{(n)})={\cal S}_4^{(n)}, }\label{PINOK}
\end{eqnarray}
where 
\beq
{\cal S}_4^{(n)}=-\sum_{p=1}^{n-1}[(\widehat{\cal T}_4^{(n-p)}+\widehat{\cal W}_4^{(n-p)})\Psi_4^{(p)}+\widehat{\cal P}_3^{(n-p)}\Psi_3^{(p)}], \qquad(n \geq 2)\,\,.
\eeq
In the case of a type D geometry $\widehat{\cal W}_4^{(0)}=\widehat{\cal W}_4^{(1)}=\widehat{\cal P}_3^{(0)}=0$  and the set of equations (\ref{PINOK}) becomes:
\begin{eqnarray}
\meqalign{
&0=0,\cr
&\widehat{\cal T}_4^{(0)}\Psi_4^{(1)}=0,\cr
&\qquad\vdots\cr
&\widehat{\cal T}_4^{(0)}\Psi_4^{(n)}={\cal S}_4^{(n)} \, .
}
\label{xteuko4}
\end{eqnarray}
The previous results are an alternative but equivalent formulation  of the work of Lousto and Campanelli concerning the higher order perturbations of a type D vacuum metric. 
The same perturbation formulation can easily be written for eqs.~(\ref{DR1213SUPERCOMPACT}), (\ref{PSIQUARTO}) and (\ref{FINPSI2}) allowing a systematic formulation of the gravitational perturbation for every vacuum spacetime.
In fact by introducing the operators:
\begin{eqnarray}
\meqalign{
&\widehat{\cal A}_1=[(D-\epsilon+\epsilon^*-3\rho-\rho^*)(\Delta+2\mu-2\gamma)-(\delta-\alpha^*-\beta+\pi^*-3\tau)(\delta^*+2\pi-2\alpha)+6\Psi_2],\cr
&\widehat{\cal B}_1=2(\kappa\nu-\sigma\lambda),\cr
&\widehat{ K}_0=[\lambda\delta-\nu D+\delta\lambda-D\nu-3\tau\lambda+\nu\epsilon-\alpha^*\lambda+3\rho\nu-\nu\epsilon^*+\lambda\pi^*+\nu\rho^*-\beta\lambda-2\Psi_3],\cr
&\widehat{ Q}_3=3[\kappa\delta-\sigma D+2\sigma\rho+2\kappa\beta-2\kappa\tau-2\sigma\epsilon],\cr
&\widehat{ M}_2=3[\Delta\kappa-\delta^*\sigma-2\mu\kappa+2\sigma\pi-\sigma\beta^*+\kappa\mu^*-3\kappa\gamma+\sigma\tau^*-\kappa\gamma^*+3\alpha\sigma],
}\end{eqnarray} 
eq.~(\ref{DR1213SUPERCOMPACT}) can be rewritten as:
\beq
(\widehat{\cal A}_1+\widehat{\cal B}_1)\Psi_1+\widehat{\cal K}_0\Psi_0+\widehat{\cal Q}_3\Psi_3+\widehat{\cal M}_2\Psi_2=0\,\, ,\label{AABB}
\eeq
and one can develop a hierarchy of perturbation equations, analogously to what we did above. 
In the case of the first order perturbation of a type D background one finds
\beq
\widehat{\cal B}_1^{(0)}=\widehat{\cal B}_1^{(1)}=\widehat{\cal K}_0^{(0)}=\widehat{\cal Q}_3^{(0)}=\widehat{\cal M}_2^{(0)}=0,
\eeq
so that the linearization of eq.~(\ref{AABB}) gives
\beq
\widehat{\cal A}_1^{(0)}\Psi_1^{(1)}+\widehat{\cal M}_2^{(1)}\Psi_2^{(0)}=0 ,
\eeq
which unfortunately couples  Weyl scalars and perturbed spin coefficents.
Relations for $\Psi_3$ can be worked out in exactly the same way. 
For $\Psi_2$  the following operators are introduced:
\begin{eqnarray}\meqalign{
&\widehat{ \cal F}_2=[(D+\epsilon+\epsilon^*-2\rho-\rho^*)(\Delta+3\mu)-(\delta+\pi^*-2\tau+\beta-\alpha^*)(\delta^*+3\pi)],\cr
&\widehat{\cal G}_2=(3\lambda\sigma-3\nu\kappa),\cr
&\widehat{ \cal N}_1=2[-\nu D+\lambda\delta-D\nu+\delta\lambda+\nu\rho^*+\lambda\pi^*-2\tau\lambda-\lambda\alpha^*+\lambda\beta-\nu\epsilon^*-\nu\epsilon+2\rho\nu+\Psi_3],\cr
&\widehat{ \cal L}_3=2[\kappa\Delta-\sigma\delta^*+\Delta\kappa-\delta^*\sigma+\alpha\sigma+\kappa\mu^*+2\mu\kappa+\tau^*\sigma-\kappa\gamma^*-\gamma\kappa-2\pi\sigma-\sigma\beta^*];\cr
}\end{eqnarray} 
eq.~(\ref{FINPSI2}) can then be rewritten as:
\beq
(\widehat{\cal F}_2+\widehat{\cal G}_2)\Psi_2+\widehat{\cal N}_1\Psi_1+\widehat{\cal L}_3\Psi_3-\Psi_0\Psi_4=0\,\,.\label{NNBB}
\eeq
Again the perturbation theory can be directly developed. In particular, for the type D case
\beq
\widehat{\cal G}_2^{(0)}=\widehat{\cal G}_2^{(1)}=\widehat{\cal N}_1^{(0)}=\widehat{\cal L}_3^{(0)}=0\, ,
\eeq
and the first order perturbations become:
\beq
\widehat{\cal F}_2^{(0)}\Psi_2^{(1)}+(\widehat{\cal F}_2^{(1)}+\widehat{\cal G}_2^{(1)})\Psi_2^{(0)}=0\,\,.\label{EEBB}
\eeq

Thus, even in the very special case of a type D geometry, the perturbation equations for $\Psi_1$,$\Psi_2$ and  $\Psi_3$ are 
coupled---order by order---with other perturbation quantities. 
This is due to the presence, in the exact equations, of terms involving $\Psi_2$  
which are responsible for the appearance of the perturbations of the spin coefficents and of the directional derivatives.
To obtain decoupled equations for these three Weyl scalars in the perturbation regime, in principle, one has to define linear operators with derivatives of order higher than the second.

We conclude this section by pointing out that our perturbation relations (\ref{xteuko1}) and (\ref{xteuko4}) assume a more compact form if compared with those found by Lousto and Campanelli  because of the completely different approach and the number of simplifying operations (always explicitly specified) that we perform at the level of the exact theory.
They use `background operators' acting on the exact Bianchi identities and then linearize the results order by order (see also their remark after eq.~(6) in \cite{LoustoCampanelli}), following the Teukolsky prescription.
The way of using exact `ad hoc' operators acting on the exact Bianchi identities and then linearizing the results was introduced by Stewart and Walker but only for first order perturbations. 

By extending such a prescription to any order
one obtains our relations.
The de Rham Laplacian mediates this task at the level of the exact theory, also explaining the nature (i.e. exact `wave equations in curved spacetime' for some of the NP Weyl and Maxwell scalars) of the relations found algebraically by Stewart and Walker.

\section{Vacuum spacetimes: test electromagnetic fields}

In this section we study the electromagnetic field equation in the absence of sources, $J^\mu=0.$ This choice has to be related to the ``exact" (in the sense of the field equations) nature of our approach, which goes beyond the perturbation level and would be corrupted by the introduction of a source term without the automatic appearance of a corresponding term in the Riemann tensor. On the other hand, eq.~(\ref{DERAEM}) below is itself an approximation, because the exact field equations should include the Riemann and Ricci terms created by the EM field and should be coupled to the gravitational de Rham Laplacian.
The formulation presented here is based on the hypothesis of test fields (no backreaction) and the goal of this analysis is to recover the existing perturbation theory, including the work of Teukolsky, Fackerell and Ipser and many other authors.

The sourcefree Maxwell equations $\delta F = 0 = dF$ for the Maxwell 2-form $F$ 
(representable as 4 complex equations for the associated NP scalars $\phi_0$, $\phi_1$, $\phi_2$, namely eqs.~(330)--(333) of Chapter 1 of \cite{chandra})  
immediately imply the vanishing of the de Rham Laplacian of the Maxwell 2-form. In a vacuum spacetime, this takes the form
\beq
[\Delta_{\rm (dR)}F]_{ab}=F_{ab\vert c}{}^{\vert c}+C_{abcd}F^{cd}=0\,, \,\,\,\label{DERAEM}
\eeq
Expressing these equations in the Newman-Penrose formalism,
one has the following wave equations.

$\bullet\qquad$ \fbox{$[\Delta_{\rm (dR)}F]_{13}=0 $}\hspace{1cm} {\it (wave equation for $\phi_0$)} 

Studying the component``13" of (\ref{DERAEM}) we get the equation for $\phi_0$:
\begin{eqnarray}
\meqalign{
&[D\Delta+\Delta D-\delta^* \delta-\delta\delta^*+(-5\gamma-\gamma^*+\mu+\mu^*)D+(-3\epsilon+\epsilon^*-\rho^*-\rho)\Delta \cr
&+(-\beta^*-\pi+5\alpha+\tau^*)\delta+(-\pi^*+\tau+3\beta+\alpha^*)\delta^*+2(\delta^*\beta+\delta\alpha-D\gamma-\Delta\epsilon\cr
&+\gamma^*\epsilon-\gamma\epsilon^*-\tau\pi+\gamma\rho^*+\pi^*\alpha+4\gamma\epsilon-\beta\tau^*+\rho\mu+\beta\beta^*-\epsilon\mu^*-\alpha\alpha^*-4\alpha\beta\cr
&+\rho\gamma-\kappa\nu+\beta\pi+\sigma\lambda-\tau\alpha-\mu\epsilon-\Psi_2]\phi_0+[4(\tau D+\kappa\Delta-\rho\delta-\sigma\delta^*)\cr
&+2(D\tau+\Delta\kappa-\delta\rho-\delta^*\sigma+\epsilon^*\tau-\rho\pi^*-\epsilon\tau+\sigma\tau^*-\kappa\gamma^*-\tau\rho^*-\sigma\beta^*+\rho\alpha^*\cr
&+\mu\kappa+\mu^*\kappa+\beta\rho-\sigma\pi-3\kappa\gamma+3\alpha\sigma+2\Psi_1)]\phi_1+[4\kappa\tau-4\sigma\rho-2\Psi_0]\phi_2=0
\,\,.}\label{EM13START}
\end{eqnarray}
First we use the commutation rules 
(\ref{comrel}); 
the next step is to divide the resulting equation by 2 and to use the two Maxwell equations:
\begin{eqnarray}\label{MAXW1}
(D-2\rho)\phi_1-(\delta^*+\pi-2\alpha)\phi_0+\kappa\phi_2=0, & \\
\label{MAXW3}
(\delta-2\tau)\phi_1-(\Delta+\mu-2\gamma)\phi_0+\sigma\phi_2=0, &
\end{eqnarray} 
to eliminate two of the directional derivatives of $\phi_1$.
Then using four Ricci identities:  
\begin{eqnarray}\meqalign{
[1/2(R_{1212}-R_{3412})]:\qquad &\Delta\epsilon=D\gamma-\alpha(\tau+\pi^*)-\beta(\tau^*+\pi)+\gamma(\epsilon+\epsilon^*)\cr
&
+\epsilon(\gamma+\gamma^*)-\tau\pi+\nu\kappa-\Psi_2, \cr
[1/2(R_{1234}-R_{3434})]:\qquad &\delta^*\beta=\delta\alpha-(\mu\rho-\lambda\sigma)-\alpha\alpha^*-\beta\beta^*+2\alpha\beta\cr
&-\gamma(\rho-\rho^*)-\epsilon(\mu-\mu^*)+\Psi_2, \cr
}\end{eqnarray}
together with the two in (\ref{RICO13}),
rewriting the Ricci identity $[R_{2431}]$  as in (\ref{RICOK2}),
multiplying it by $\phi_0$ and adding the resulting quantity (which is actually zero) to the wave equation one obtains:
\begin{eqnarray}
\meqalign{
&[(D-\epsilon+\epsilon^*-2\rho-\rho^*)(\Delta+\mu-2\gamma)-(\delta-\beta-\alpha^*-2\tau+\pi^*)(\delta^*+\pi-2\alpha)]\phi_0\cr
&+(\sigma\lambda-\kappa\nu)\phi_0+2[\kappa\Delta-\sigma\delta^*+\Delta\kappa-\delta^*\sigma-\kappa\gamma^*-\sigma\beta^*+\kappa\mu^*+\sigma\tau^*+3\alpha\sigma\cr
&-3\kappa\gamma+2\Psi_1]\phi_1-\Psi_0\phi_2=0\,, \label{EMMY1}
}\end{eqnarray}
where in the first line it is easy to recognize the operator found by Teukolsky for the electromagnetic perturbations of a type D metric.
One can also verify that (\ref{EMMY1}) is the electromagnetic equation of Stewart and Walker in the GHP formalism:
\begin{eqnarray}
\meqalign{
&[\Ph \Ph'-\Ed \Ed'-\rho'\Ph-(2\rho+\bar\rho)\Ph '+\tau'\Ed+(2\tau +\bar\tau')\Ed'+2\rho\rho'-2\tau\tau'+\Psi_2]\phi_0\cr
&+[2\Ph'\kappa-2\Ed'\sigma-2\kappa(\bar\rho'-2\rho')+2\sigma(\bar\tau-2\tau')+4\Psi_1]\phi_1+[2\kappa\Ed-2\sigma\Ph\cr
&+2\rho\sigma-2\kappa\tau-\Psi_0]\phi_2=0\, .}\label{GHPPHI0}
\end{eqnarray}
In fact ``translating" (\ref{GHPPHI0}) into the standard Newman-Penrose form one gets:
\begin{eqnarray}
\meqalign{
&[(D-\epsilon+\epsilon^*)(\Delta-2\gamma)-(\delta-\beta-\alpha^*)(\delta^*-2\alpha)+\mu(D-2\epsilon)\cr
&-(2\rho+\rho^*)(\Delta-2\gamma)-\pi(\delta-2\beta)+(2\tau-\pi^*)(\delta^*-2\alpha)-2\rho\mu+2\tau\pi+\Psi_2]\phi_0\cr
&+2[(\Delta-3\gamma-\gamma^*)\kappa-(\delta^*-3\alpha+\beta^*)\sigma+\kappa(\mu^*-2\mu)+\sigma(\tau^*+2\pi)+2\Psi_1]\phi_1\cr
&+[2\kappa(\delta+2\beta)-2\sigma(D+2\epsilon)+2\rho\sigma-2\kappa\tau-\Psi_0]\phi_2=0 \,.
}\label{TRANSLGHPPHI0}\end{eqnarray}
Using the following two Maxwell equations:
\begin{eqnarray}
\meqalign{
&(D-\rho+2\epsilon)\phi_2-(\delta^*+2\pi)\phi_1+\lambda\phi_0=0,\cr
&(\delta-\tau+2\beta)\phi_2-(\Delta+2\mu)\phi_1+\nu\phi_0=0, \cr
}
\end{eqnarray}
in the previous equation and adding the Ricci identity $[R_{2431}]$ 
(see eq.~\ref{RICOK2}) multiplied by $\phi_0$, one finds identically eq.~(\ref{EMMY1}).

$\bullet\qquad$ \fbox{$[\Delta_{\rm (dR)}F]_{24}=0 $}\hspace{1cm} {\it (wave equation for $\phi_2$)} 

The equation for $\phi_2$ can be immediately obtained form eq.~(\ref{EMMY1}). Using the exchange symmetry ${\bf l}\leftrightarrow {\bf n}$, ${\bf 
m}\leftrightarrow {\bf m^{*}}$ of the Newman-Penrose quantities
and the relation (\ref{SCAMMAX}), one gets

\begin{eqnarray}
\meqalign{
&-\{[(\Delta+\gamma-\gamma^*+2\mu+\mu^*)(D-\rho+2\epsilon)-(\delta^*+\alpha+\beta^*+2\pi-\tau^*)(\delta-\tau+2\beta)]\phi_2\cr
&+(\sigma\lambda-\kappa\nu)\phi_2+2[\lambda\delta-\nu D+\delta\lambda-D\nu-\nu\epsilon^*-\lambda\alpha^*+\nu\rho^*+\lambda\pi^*+3\beta\lambda\cr
&-3\nu\epsilon+2\Psi_3]\phi_1-\Psi_4\phi_0\}=0\,\,\,.\label{EMMY2} 
}\end{eqnarray}
The overall minus factor in our case can be eliminated but this form is convenient for further developments in the presence of sources.
The GHP analogue of this equation is

\begin{eqnarray}
\meqalign{
&[\Ph' \Ph-\Ed' \Ed-\rho\Ph '-(2\rho '+\bar\rho ')\Ph+\tau\Ed'+(2\tau'+\bar\tau)
\Ed+2\rho'\rho-2\tau'\tau+\Psi_2]\phi_2\cr
&+[2\Ph\kappa'-2\Ed\sigma'-2\kappa'(\bar\rho-2\rho)+2\sigma'(\bar\tau '-2\tau)+4\Psi_3]\phi_1
+[2\kappa'\Ed'-2\sigma'\Ph'\cr
&+2\rho'\sigma'-2\kappa'\tau'-\Psi_4]\phi_0=0.
}\label{GHPPHI2}
\end{eqnarray}

$\bullet\qquad$ \fbox{$[\Delta_{\rm (dR)}F]_{12}-[\Delta_{\rm (dR)}F]_{34}=0 $}\hspace{1cm} {(\it wave equation for $\phi_1$)}

To obtain the equation for $\phi_1$ one has to subtract the components ``12" and ``34" of (\ref{DERAEM}) in order to eliminate all the $\phi_n^*$ terms.
The resulting wave equation, divided by a factor of 2, is: 
\begin{eqnarray}
\meqalign{
&[D\Delta+\Delta D-\delta^* \delta-\delta\delta^*+(-\gamma-\gamma^*+\mu+\mu^*)D+(\epsilon+\epsilon^*-\rho^*-\rho)\Delta \cr
&+(-\beta^*-\pi+\alpha+\tau^*)\delta+(-\pi^*+\tau-\beta+\alpha^*)\delta^*+4(\rho\mu+\lambda\sigma-\tau\pi-\nu\kappa+\Psi_2)]\phi_1\cr
&[-2\nu D-2\pi\Delta+2\lambda\delta+2\mu\delta^*-D\nu-\Delta\pi+\delta\lambda+\delta^*\mu-\lambda\alpha^*+\rho\nu+\pi\gamma^*-\mu\tau^*\cr
&-\tau\lambda-3\alpha\mu+\lambda\pi^*-\pi\mu^*+\mu\beta^*-\lambda\beta-\nu\epsilon^*+\nu\rho^*+3\gamma\pi+\nu\epsilon-2\Psi_3]\phi_0\cr
&+[2\tau D+2\kappa\Delta-2\rho\delta-2\sigma\delta^*+D\tau+\Delta\kappa-\delta\rho-\delta^*\sigma-\sigma\pi-\tau\rho^*+\gamma\kappa-\kappa\gamma^*\cr
&+3\tau\epsilon+\tau^*\sigma-3\rho\beta+\mu^*\kappa-\alpha\sigma+\tau\epsilon^*-\sigma\beta^*+\rho\alpha^*+\mu\kappa-\rho\pi^*-2\Psi_1]\phi_2=0.
}\label{EMPSI1START}
\end{eqnarray}
The procedure to simplify the previous relations is the usual one.
The starting point is to rewrite the second order part by using (\ref{eq:dirder}); dividing the resulting equation by 2 and using all four NP Maxwell equations,
some of the directional derivatives of $\phi_0$ and $\phi_2$ are eliminated in favor of the derivatives of $\phi_1$. 
Now, using the Ricci identies: (\ref{RICOK1}), (\ref{RICO13}) and
\begin{eqnarray}\meqalign{
[R_{2443}]:\qquad &\delta\lambda-\delta^*\mu=\nu(\rho-\rho^*)+\pi(\mu-\mu^*)+\mu(\alpha+\beta^*)+\lambda(\alpha^*-3\beta)-\Psi_3\,, \cr
[R_{2421}]:\qquad &D\nu-\Delta\pi=\mu(\pi+\tau^*)+\lambda(\pi^*+\tau)+\pi(\gamma-\gamma^*)-\nu(3\epsilon+\epsilon^*)+\Psi_3\,;
}\end{eqnarray}
adding the Ricci identity (\ref{RICOK2}) multiplied by $2\phi_1$, we finally obtain
\begin{eqnarray}
\meqalign{
&[(D+\epsilon+\epsilon^*-\rho-\rho^*)(\Delta+2\mu)-(\delta+\beta-\alpha^*-\tau+\pi^*)(\delta^*+2\pi)]\phi_1\cr
&+[-\nu D+\lambda\delta-D\nu+\delta\lambda-\lambda\alpha^*-\epsilon^*\nu+\rho\nu-\tau\lambda+\lambda\beta-\nu\epsilon+\nu\rho^*+\pi^*\lambda]\phi_0\cr
&+[\kappa\Delta-\sigma\delta^*+\Delta\kappa-\delta^*\sigma-\sigma\beta^*-\kappa\gamma^*+\mu\kappa-\sigma\pi+\kappa\mu^*+\sigma\tau^*+\alpha\sigma-\kappa\gamma]\phi_2=0\,.
}\label{AAXXWW}
\end{eqnarray}

It is easily seen that in the (exact) eq.~(\ref{AAXXWW}) the $\phi_1$ part is exactly the Newman-Penrose translation of the GHP equation which describes the $\phi_1$ electromagnetic perturbation in a vacuum type D metric present in the paper of Lun and Fernandez \cite{Fernandez}:
\begin{equation}
[(\Ph-\rho-\bar\rho)(\Ph'+2\mu)-(\Ed-\tau+\bar\pi)(\Ed'+2\pi)]\phi_1=0\,,
\end{equation}
which, after a tetrad transformation which sets $\epsilon=0$,  becomes the well known Fackerell and Ipser equation \cite{FackerellIpser}.

It is important to get the general vacuum equation for $\phi_1$ in the GHP formalism, which can be  quite easily obtanined by following the Stewart and Walker prescriptions. We only give the final result:
\begin{equation}
\meqalign{
&[(\Ph-\rho-\bar\rho)(\Ph'-2\rho')-(\Ed-\tau-\bar\tau')(\Ed'-2\tau')]\phi_1\cr
&+[\kappa' \Ph+\Ph \kappa'-\sigma' \Ed-\Ed \sigma'+\sigma'\tau+\sigma'\bar\tau'-\kappa'\rho-\kappa'\bar\rho  ]\phi_0\cr
&+[\kappa \Ph'+\Ph'\kappa-\sigma\Ed'-\Ed'\sigma+\bar\tau\sigma+\tau'\sigma-\rho'\kappa-\bar\rho'\kappa]\phi_2=0\,\,.\label{ECEDX}
}
\end{equation}
Translating (\ref{ECEDX}) into standard Newman-Penrose formalism one gets (\ref{AAXXWW}). 

At this point the (first order) perturbation analysis can be performed. 
In the type D case, recalling that for vacuum spacetimes one has to use the test field approximation (no changes in the vacuum background means no changes in the tetrad vectors,
spin coefficients and Weyl scalars), it is straightforward to show the coincidence of the linearized form of (\ref{EMMY1}), (\ref{EMMY2}) and (\ref{AAXXWW}) with the Teukolsky, Fackerell-Ipser and Lun-Fernandez equations.
In a more general vacuum geometry the decoupling is not present and one obviously recovers  the Stewart and Walker theorems. 

\section{Concluding remarks}
In this article we have presented a new form of the Teukolsky Master Equation and a new point of view for perturbations in General Relativity, starting from exact wave equations in curved vacuum spacetimes valid for the Riemann and Maxwell tensors themselves and involving both the generalized and the ordinary de Rham Laplacian respectively.
Once perturbed these equations immediately give the standard perturbation theory of Teukolsky, Stewart and Walker, Lousto and Campanelli, explaining also the true origin and the geometrical meaning of the same Teukolsky Master Equation.
We have discussed our exact wave equations both in the Newman-Penrose and Geroch-Held-Penrose formalisms, collecting most of results scattered in literature in a series of papers and books and over a period of more than thirty years.
We have also presented a systematic study of perturbation theory at every order, which recovers and extends existing results on this subject. 
Furthermore, our approach can be directly generalized to nonvacuum spacetimes:1)
The de Rham Laplacian of the Riemann tensor then involves (in the exact regime) 
the Ricci tensor and curvature scalar terms, in turn to be re-expressed in terms of energy-momentum tensor
of the sources or directly coupled to the de Rham equations for the other corresponding physical fields: Maxwell, Yang Mills, Dirac, Rarita-Schwinger\cite{PenroseRindler}; 2) 
the perturbation regime can be developed straightforwardly, recovering,  for instance, once our general results are specialized to a vacuum type D background, the ``matter perturbed sources" appearing in the works of Teukolsky and of Lousto and Campanelli. 

Regarding the tetrad and gauge invariance of the theory, the procedure that one must follow is the standard one, 
well discussed in the literature by Bruni, Matarrese, Mollerach and Sonego\cite{bruni1,bruni2} and by \cite{LoustoCampanelli} 
and we will adapt it to this formalism in a forthcoming paper. 

We conclude pointing out that our approach could have useful consequences for the still open problem of the electromagnetic and gravitational coupled perturbations of the Kerr-Newman spacetime\cite{Lee1,Chitre,Lee2,Fackerell} as well as for the more general problem of propagation of gravitational waves on any given background.

\end{document}